\documentclass[manuscript,screen,nonacm]{acmart}

\AtBeginDocument{%
  }
    
\usepackage{cleveref}
\usepackage{graphicx}
\usepackage{subcaption}
\usepackage{listings}
\usepackage{xcolor}
\usepackage{framed}

\definecolor{bg}{gray}{.99}
\definecolor{codeblue}{rgb}{0,0,0.8}
\definecolor{codegreen}{rgb}{0.25,0.49,0.48}
\definecolor{codegray}{rgb}{0.5,0.5,0.5}

\lstdefinestyle{pythonbox}{
  language=Python,
  backgroundcolor=\color{bg},
  basicstyle=\fontsize{7}{9}\selectfont\ttfamily,
  keywordstyle=\color{codeblue}\bfseries,
  stringstyle=\color{codegreen},
  commentstyle=\color{codegray}\itshape,
  numberstyle=\fontsize{6}{8}\selectfont\ttfamily\color{codegray},
  numbers=left,
  numbersep=8pt,
  breaklines=true,
  keepspaces=true,
  showstringspaces=false,
  tabsize=4,
  xleftmargin=20pt,
  frame=none,
  lineskip=-1pt,
}

\begin{document}
\title{XARP: A Human-First and Agent-Ready Extended Reality Toolkit in Python}
\author{Arthur Caetano}
\email{caetano@ucsb.edu}
\affiliation{  \institution{University of California, Santa Barbara}
  \city{Santa Barbara}
  \state{CA}
  \country{USA}
}
\author{Radha Kumaran}
\email{kumaran@cs.uni-saarland.de}
\affiliation{  \institution{Saarland University}
  \city{Saarbrücken}
  \state{Saarland}
  \country{Germany}
}
\author{Kelvin Jou}
\email{kelvinjou@ucsb.edu}
\affiliation{  \institution{University of California, Santa Barbara}
  \city{Santa Barbara}
  \state{CA}
  \country{USA}
}
\author{Tobias Höllerer}
\email{holl@ucsb.edu}
\affiliation{  \institution{University of California, Santa Barbara}
  \city{Santa Barbara}
  \state{CA}
  \country{USA}
}
\author{Misha Sra}
\email{sra@ucsb.edu}
\affiliation{  \institution{University of California, Santa Barbara}
  \city{Santa Barbara}
  \state{CA}
  \country{USA}
}
\renewcommand{\shortauthors}{Caetano et al.}

\begin{abstract}

Building XR-AI research prototypes requires navigating two largely separate ecosystems. Mainstream XR development relies on C\#/C++ and game engines, while AI development is centered on Python. This toolchain fragmentation slows down contributions to human-AI spatial interaction research. To broaden access to XR development in the Python ecosystem, we present XARP (XR Agent-ready Remote Procedures), a toolkit for rapid XR-AI prototyping in Python. XARP application logic runs on a Python server and controls a Unity client through WebSocket messages.
This architecture enables compatibility with multiple client platforms and live reloading of application code without client redeployment.
XARP is available to humans as a library and to AI agents as callable tools and through Model Context Protocol.
We designed XARP through formative case studies and refined it through an early acceptance evaluation with 24 XR and AI developers and a six-week longitudinal study with two developers building an independent research project.
Potential users expected the toolkit to improve their performance and facilitate development. Sustained use confirmed faster iteration and easier setup compared to conventional XR workflows, with asset-intensive and performance-critical projects emerging as the clearest limitations. Technical benchmarks show that hand and head tracking data streaming was close to the device refresh rate of 72 FPS, and that AI agents using XARP consumed 19\% fewer tokens than those writing equivalent C\# Unity code. Beyond broadening access to XR development, XARP reduces engineering friction in spatial computing research and opens new pathways for AI agents to participate in XR application development.
XARP is open source and available at \url{https://github.com/hal-ucsb/xarp}.
\end{abstract}

\setcopyright{cc}
\setcctype{by}
\acmJournal{PACMHCI}
\acmYear{2026} \acmVolume{10} \acmNumber{4} \acmArticle{EICS010}
\acmMonth{6} \acmDOI{10.1145/3816762}

\maketitle

\begin{figure}[t]
\centering
\includegraphics[width=\linewidth]{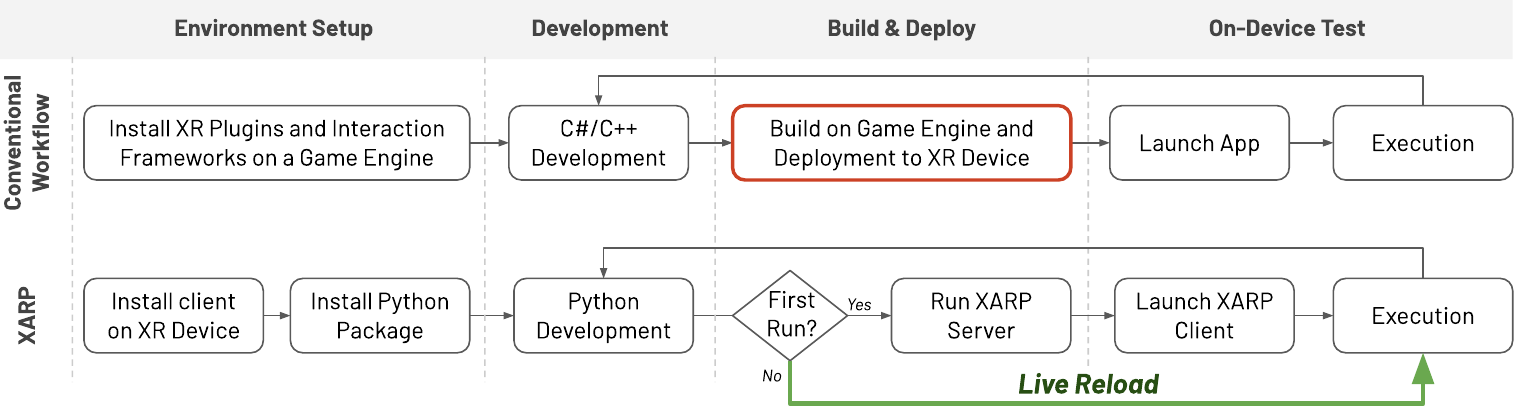}
\caption{Setting up XARP requires only installing a Python package and the client binary on the XR device, avoiding game engines. Development leverages tools and practices already familiar to Python developers. XARP offers live reloading, which allows app modifications to take effect on XR devices in a few seconds without redeployment.}
\label{fig:workflow}
\Description[XARP avoids build-redeploy cycles with live reload.]{Workflow diagram comparing conventional XR development with XARP. Conventional development requires installing XR plugins in a game engine, writing C\#/C++ code, building and deploying to the XR device, launching the app, and executing it. XARP requires installing a client on the XR device and a Python package, writing Python code, starting the XARP server only on the first run, launching the XARP client, and executing the app. A green live-reload loop shows that later changes can run on the device without rebuilding or redeploying.}
\end{figure}

\section{Introduction}
\label{sec:intro}

Human-AI spatial interaction has a long research trajectory, from early work supporting aircraft piloting~\cite{furness1986super} and machinery maintenance~\cite{feiner1993knowledge} to recent advances in context-aware guidance~\cite{srinidhi2024xair, li2025satori, Zhao2025GuidedRG, xu2024multimodal} and scalable authoring~\cite{giunchi2024dreamcodevr, de2024llmr, zhu2025agentar}. Despite this long history and renewed interest, building XR-AI research prototypes remains a practical challenge. The XR development ecosystem relies on C\#/C++ frameworks and game engines~\cite{borsting2022agenda, frau2023xrspotlight, bose2024empirical} and is largely separated from the Python ecosystem that dominates AI development~\cite{maslej2025aiindex, stackoverflow2025devsurvey}. While XR developers can integrate AI via web APIs or on-device inference\footnote{Unity Inference Engine: \url{https://docs.unity3d.com/Packages/com.unity.sentis@2.1/manual/index.html}, Accessed March 27, 2026.}, AI developers using Python have limited options for developing XR interfaces~\cite{pyopenxr, py4godot, godotpythonextension}. This asymmetrical gap hinders Python developers, the majority of AI researchers~\cite{maslej2025aiindex}, from contributing to research in human-AI spatial interaction, delaying the benefits these applications may provide to society.

To close this gap, we present XARP (\textbf{X}R \textbf{A}gent-ready \textbf{R}emote \textbf{P}rocedures), a toolkit for rapid XR-AI prototyping in Python. With XARP, Python developers can quickly build XR interfaces in their familiar ecosystem, enabling a workflow similar to what Streamlit\footnote{Streamlit: \url{https://streamlit.io}. Accessed April 3, 2026.} and Gradio\footnote{Gradio: \url{https://www.gradio.app}. Accessed April 3, 2026.} offer for web interfaces. Our toolkit provides high-level abstractions for common XR operations, including displaying 3D meshes and hand tracking.
Internally, XARP implements a client-server architecture in which application logic executes on a Python server that controls a Unity client through commands encoded in  MessagePack, an efficient binary alternative to JSON, transmitted over a WebSocket connection.

This design has several benefits.
First, during development, changes to application logic on the server take effect on the XR device without redeploying the client. This live reloading feature bypasses the lengthy build-deploy cycles that slow down iteration and break creative flow in XR development.
Second, the client abstracts device-specific details from XARP applications, making them portable to any XR platform with a compatible client. A single port of the Unity client to a new platform is sufficient to make all XARP applications compatible with it.
Third, AI agents can use XARP as a set of callable tools or through a built-in Model Context Protocol (MCP) server~\cite{mcp}, enabling them to control XR environments at the same abstraction level as human developers.

Our design and evaluation methodology consisted of four stages.
In a formative phase, we conducted two case studies with novice XR developers to elicit design requirements from their experiences and develop an initial version of the toolkit.
Next, we conducted an early acceptance evaluation through an online video walkthrough and survey with 24 XR and AI developers.
In a longitudinal study, two XR-AI developers used XARP over six weeks to develop a prototype for an independent research project.
Finally, we performed technical evaluations comparing streaming performance across different configurations and AI agent token consumption in code generation tasks using XARP versus Unity C\#.

We demonstrate that XARP effectively enables XR interface development in Python through ``how-to'' scenarios with example applications~\cite{ledo2018evaluation}, showcasing live reloading and multiplatform capabilities.
Our human evaluations show that XR novices familiar with Python benefit from the lower entry~\emph{threshold}~\cite{myers2000past}, while experienced XR developers benefit from fast prototyping.
They further delineate the toolkit's \emph{ceiling}~\cite{myers2000past}, distinguishing future development roadmap items (e.g., trackable physical elements) from structural limitations (e.g., fine-grained graphics control).
Technical evaluation shows that agents using XARP consume fewer tokens at the median than their C\#/Unity counterparts. Hand and head streaming performance approached the device refresh rate of 72 FPS under ideal conditions.

Beyond enabling XR development for new audiences, XARP advances human-AI spatial interaction research by removing \emph{accidental complexity}~\cite{brooks1987silver,fraser2007reloaded}, allowing researchers to concentrate on \emph{essential} aspects. A shared abstraction for XR accelerates hypothesis testing and improves reproducibility through controlled cross-platform studies and systematic comparison of design alternatives. More broadly, such abstractions become increasingly critical as AI agents take on more active roles in software development and academic research~\cite{jimenez2024swebench,jowsey2025reject}.
This article is organized in two parts. \Crefrange{sec:how_to}{sec:arch} offer hands-on demonstrations and a toolkit description; \cref{sec:rw} onwards presents related work, methodology, evaluations, and findings.

\begin{figure}[t]
\centering
\begin{minipage}[t]{0.51\textwidth}
  \vspace{0pt}
  \begin{lstlisting}[style=pythonbox]
from xarp.server import run, show_qrcode_link

def hello_world(xarp, params):
    # Get connection parameters
    hi = f'Hello, {params["name"]}!' 
    xarp.write(hi, title="XARP")

if __name__ == '__main__':
    # QR code with connection parameters
    show_qrcode_link(name="world") 
    
    # run the application entrypoint
    run(hello_world)\end{lstlisting}
\end{minipage}
\begin{minipage}[t]{0.45\textwidth}
  \vspace{0pt}
  \centering
  \includegraphics[width=\linewidth]{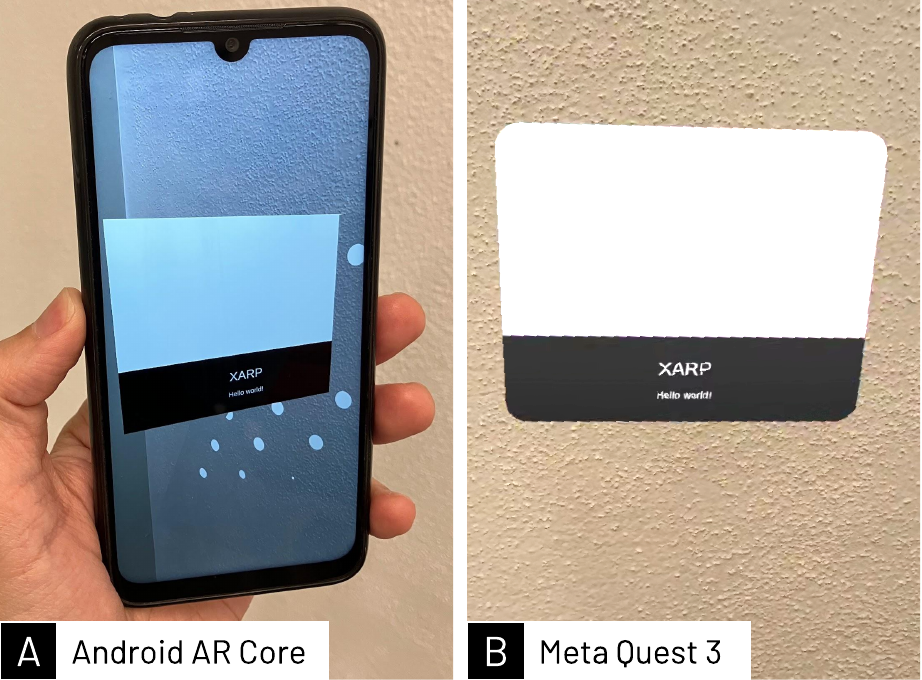}
\end{minipage}
\caption{XARP enables XR development in Python. The same ``hello world'' application running on (a) Android ARCore and (b) Meta Quest 3, showcasing XARP's multiplatform capabilities enabled by client-server decoupling.}
\label{fig:hello_world}
\Description[XARP is multiplatform]{Figure showing a Python-based XARP ``hello world'' XR application and its execution on two platforms. On the left is Python code importing functions from xarp.server, defining a hello\_world function that writes a greeting to the XR display, generating a QR code link with connection parameters, and running the application. On the right are two photographs demonstrating the same application running on different devices: (A) an Android ARCore smartphone displaying a floating white panel labeled ``XARP'' with ``Hello world!'' in an augmented reality scene, and (B) a Meta Quest 3 headset view showing the same floating panel attached to a wall. The figure illustrates XARP’s cross-platform XR capability through a client-server architecture using Python.}
\end{figure}

\section{Developing with the XARP Library}
\label{sec:how_to}

This section presents a how-to scenario followed by an example application~\cite{ledo2018evaluation}, demonstrating that XARP supports a complete XR development workflow within the Python ecosystem.
The how-to scenario covers environment setup, app development, live reloading, and deployment to two platforms, demonstrating the toolkit's threshold~\cite{myers2000past}.
The 3D object reconstruction example application demonstrates that XARP can support common XR-AI research needs.

\subsection{How-To Scenario - Main Workflow}
\label{sec:how-main-workflow}

\subsubsection{Environment Setup}
To get started, developers install the XARP Python package directly from the \href{https://github.com/anonymized}{open-source repository} using a package manager. Next, they download and install a XARP client compatible with their XR device, also available in the open-source repository. This step requires no game engine and relies only on tools provided by the device vendor\footnote{Meta Quest Developer Hub: \url{https://developers.meta.com/horizon/documentation/unity/ts-mqdh}. Accessed March 27, 2026.
\par Android Debug Bridge: \url{https://developer.android.com/tools/adb}. Accessed March 27, 2026.}. Prebuilt client binaries are available for Meta Quest and Android ARCore, with the Unity client source also available for developers who wish to extend or modify it. Porting XARP to additional platforms is part of the future roadmap.

\subsubsection{XR Development in Python}
Once setup is complete, developers write their application as an entry point function and run it by calling \texttt{xarp.server.run} as seen in the code snippet in~\cref{fig:hello_world}. This function generates a QR code encoding the server's local address and any developer-defined parameters. The client scans it to connect automatically to the server and launch the app.

\subsubsection{Live Reloading}
Developers can modify the code and restart the server to see the changes take effect directly on a running client. This is possible because the client briefly disconnects when the server restarts and then automatically reconnects either to the last connected address or to a new QR code. Once the client reconnects, the new application code is executed. This live-reloading mechanism takes less than 30 seconds, compared to conventional build-and-deploy processes that commonly exceed 10 minutes on first-time deploys. Tethered execution on game engines\footnote{Unity Link: \url{https://developers.meta.com/horizon/documentation/unity/unity-link/}. Accessed April 30, 2026.} also bypasses the build step but requires a Windows machine and renders via the workstation's GPU, meaning results do not reflect real on-device performance. XARP's live reloading runs directly on the target device without tethering, preserving both iteration speed and hardware fidelity. By shortening the time from modification to test on the target platform, XARP contributes to faster iterations and allows developers to stay in a creative flow. This process is depicted in~\cref{fig:workflow}.

\subsubsection{Multiplatform Execution}
Another benefit of using XARP is that the same application can run on multiple XR platforms. The only necessary step is to install the appropriate client and scan the server's QR code; the application itself remains unchanged. This is possible because clients work as runtime abstractions that map commands received from the server into platform-specific functions. This approach leverages the OpenXR multiplatform capabilities offered through Unity while removing the burden of porting each application from end developers.~\Cref{fig:hello_world} shows the same app running on the Meta Quest 3 and on an Android device.

\begin{figure}[b]
\centering
\begin{minipage}[t]{0.52\textwidth}
    \vspace{0pt}
    \begin{lstlisting}[style=pythonbox]
# start streaming hand poses
stream = xarp.sense(hands=True) 
for frame in stream:
    hands = frame["hands"]
    # check for a right-hand pinch
    if hands.right and pinch(hands.right): 
        # capture RGB image
        img = xarp.image() 
        job_id = util.image_to_3d_job(img)
        break

# save the 3D mesh on the device
mesh_asset = GLBAsset(
    asset_key = "mesh_asset",
    raw = util.download_glb(job_id))
xarp.save(mesh_asset)

# display the 3D mesh
mesh = Element(
    key = "mesh_element",
    asset = mesh_asset)
xarp.update(mesh)

# manipulate the mesh with a hand palm
for frame in stream:
    hands = frame["hands"]
    if hand := hands.right:
        p = hand[PALM]
        mesh.transform.position = p.position
        mesh.transform.rotation = p.rotation
        xarp.update(mesh)\end{lstlisting}
\end{minipage}
\begin{minipage}[t]{0.45\textwidth}
    \vspace{0pt}
    \includegraphics[width=\linewidth]{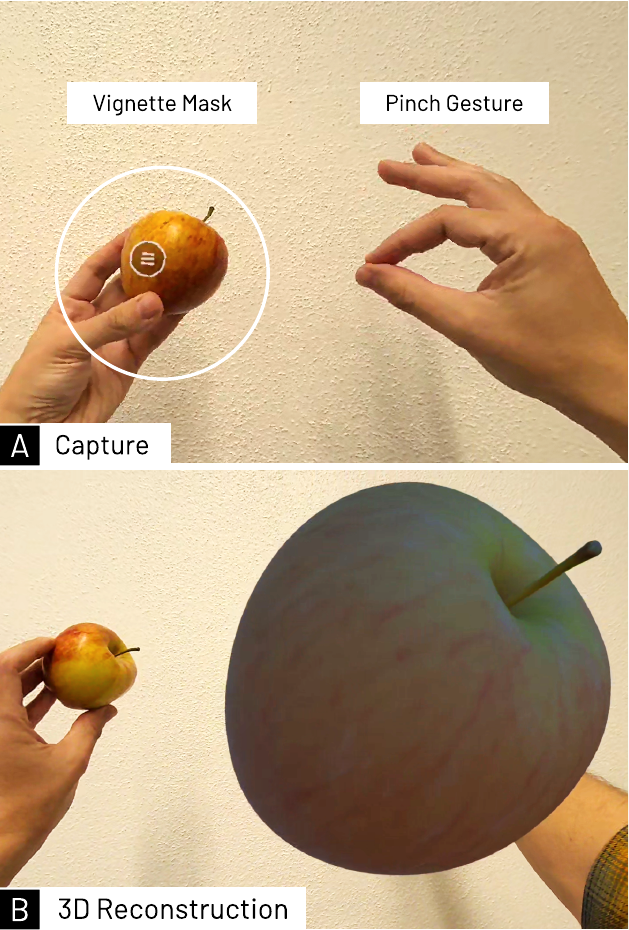}
\end{minipage}
\caption{3D object reconstruction app. (a) A pinch gesture detection (line 6) triggers an image capture with vignette masking. (b) The reconstructed object is displayed in XR and controlled via hand tracking (lines 28-30).}
\label{fig:reconstruction}
\Description{Figure illustrating a Python-based XR application for 3D object reconstruction and hand interaction using XARP. On the left is Python code that streams hand poses, detects a right-hand pinch gesture, captures an RGB image, reconstructs a 3D mesh from the image, saves and displays the mesh in XR, and updates the mesh position and rotation according to tracked hand-palm motion. On the right are two demonstration images. In panel (A), a user holds an apple while performing a pinch gesture with the other hand; a circular vignette mask highlights the object capture region used for image acquisition. In panel (B), the reconstructed 3D apple mesh appears enlarged in XR beside the real apple and follows the user’s hand movement through hand tracking.}
\end{figure}

\subsection{Example Application - 3D Object Reconstruction}
In-situ 3D object reconstruction is an active research area in human-AI spatial interaction. Systems in this category often use image-to-3D pipelines for fabrication~\cite{wang2025dreamprinting}, collection~\cite{li2025interecon}, and remote collaboration~\cite{huang2024virtualnexus}. This example demonstrates gesture-based object selection and interactive visualization that can serve as a scaffold for similar systems. Full implementation of the code snippet shown in~\cref{fig:reconstruction} can be found in the supplemental material.

\subsubsection{Physical Object Selection}
The example application streams hand poses and checks for a pinch gesture to trigger an RGB image capture. A vignette mask highlights the center of the image, and an image-to-3D pipeline~\cite{sam3dteam2025sam3d3dfyimages} generates a 3D object reconstruction in GLB format from the masked region. Gesture detection uses a deterministic joint angle test, as shown in line 6 of the code snippet in~\cref{fig:reconstruction}(a). Beyond hand tracking, XARP's streaming mechanism supports additional sensing modalities, such as depth images and head positions.

\subsubsection{Interactive Visualization}
The reconstructed object is displayed as a 3D element and manipulated using hand tracking and pinch gestures. XARP supports multiple element types, including text labels, image panels, 3D primitives, GLB objects, video, and audio.~\cref{fig:reconstruction}(b) shows 3D object reconstruction controlled via the right-hand palm (lines 28--30).

\section{Developing with XARP Agents}
\label{sec:how_to_agents}

\begin{figure}[h]
\centering

\begin{minipage}[t]{0.5\textwidth}
    \vspace{0pt}
    \begin{lstlisting}[style=pythonbox]
for frame in stream:
    hands = frame["hands"]
    if hands.left and victory(hands.left):
        if recording:
            break
        recording = True
        xarp.say("Tracking")

    # motion tracking
    if recording and hands.right:
        palm = hands.right[PALM]
        poses.append(palm)
        dumbbell.transform = Transform(
            position=palm.position,
            rotation=palm.rotation)
        xarp.update(dumbbell)

# xr agent feedback
agent.run("Analyze the palm poses during a standard dumbbell curl and find critical mistakes. Create a visual demonstration that combines labels, primitives, and colors to help users improve their technique. Make suggestions using speech.", additional_args=dict(poses=poses))\end{lstlisting}
\end{minipage}
\begin{minipage}[t]{.475\textwidth}
    \vspace{0pt}
    \includegraphics[width=\linewidth]{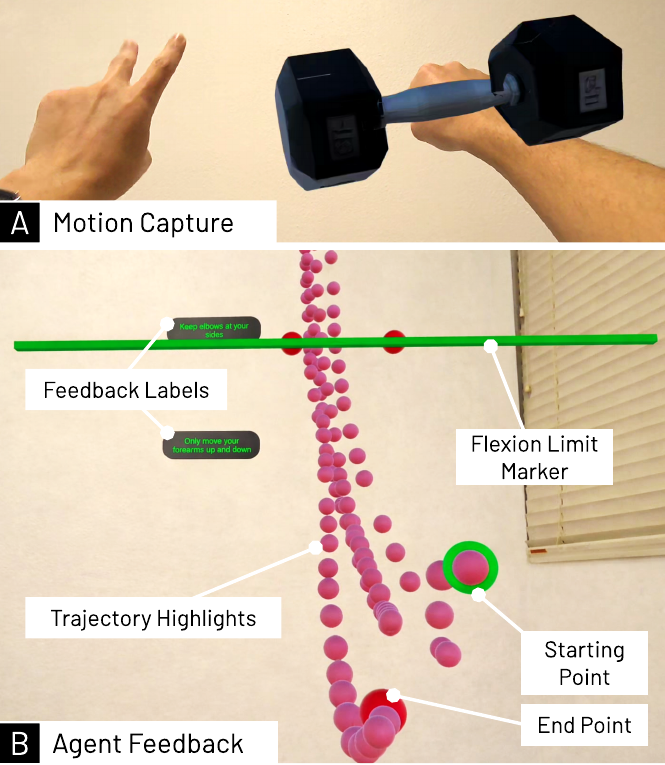}
\end{minipage}
\caption{Fitness coaching application demonstrating agent-generated feedback. (a) A V gesture initiates tracking and displays a virtual dumbbell. (b) After the exercise is complete, the agent analyzes motion data and generates multimodal feedback, in this case including motion trajectory highlight, biceps flexion limit marker, text labels with instruction and feedback, and speech output.}
\label{fig:dumbbell}
\Description{Figure showing a fitness coaching XR application with agent-generated feedback implemented in Python using XARP. On the left is Python code that streams hand-pose data, detects a left-hand victory gesture to start recording, tracks right-hand palm motion during a dumbbell curl exercise, updates the pose of a virtual dumbbell in XR, and sends recorded palm trajectories to an AI agent for analysis and multimodal feedback generation. On the right are two demonstration panels. In panel (A), a user performs a victory gesture while holding a dumbbell; this gesture initiates motion capture and displays a virtual dumbbell for tracking. In panel (B), the XR system visualizes analyzed exercise motion with pink trajectory markers, a green flexion limit line, highlighted starting and ending points, and floating feedback labels containing exercise guidance. The figure demonstrates agent-driven coaching feedback combining motion analysis, visual overlays, and speech output.}
\end{figure}

Agents are autonomous systems that perceive and act in their environment on behalf of a user~\cite{maes_shneiderman1997agents_vs_direct}. Agents can leverage vision-language models for visual understanding, natural language reasoning, and planning, and act through tool calling~\cite{wang2024executable}. XARP serves as a set of callable tools that enable agents to perceive and act in XR environments, leveraging similar abstractions to those used by human developers.

\subsection{How-To Scenario - XR Agents Workflow}

\subsubsection{Agent Injection}
To accelerate integration with AI agents, XARP offers an out-of-the-box integration with Hugging Face smolagents, a model-agnostic agentic framework\footnote{Hugging Face, \textit{smolagents}: \url{https://huggingface.co/docs/smolagents}. Accessed March 27, 2026.}. XARP is compatible with any agentic framework that can convert plain Python functions into callable tools. After following the same setup presented in~\cref{sec:how-main-workflow}, developers include a new \texttt{agent} argument in their entry point function and execute it using the \texttt{xarp.agents.run\_agent} function. XARP will automatically create an agent equipped with XR tools and inject it into the entry point function. At any point in the code, developers can delegate to agents, as seen in the code panel in~\cref{fig:dumbbell}, lines 19 and 20.

\subsubsection{Programmatic Control vs. Generative Delegation}

XARP opens a new possibility for XR developers by combining programmatic control through the library with the option of delegating application behavior to a runtime agent. Developers must decide which requirements are well-defined enough to implement programmatically and which are better delegated to an agent that can generate code dynamically to meet situated needs at runtime. This decision is shaped by the tradeoff between control and delegation studied in human-AI interaction~\cite{maes_shneiderman1997agents_vs_direct, horvitz1999principles}. 

XARP supports a full control-delegation spectrum. At one end, developers retain complete control, using XARP purely as a library with no agent involvement, as demonstrated in~\cref{sec:how_to}. In the middle, selective delegation allows developers to implement the core application flow programmatically while delegating specific, context-dependent behaviors to an agent. For instance, this could involve generating personalized visual feedback based on user input, as seen in~\cref{fig:dumbbell}. This approach can serve as a probe~\cite{martelaro2017needfinding, boehner2007hci} at the cost of introducing latency and non-deterministic behavior. At the other extreme, full delegation hands the entire application behavior to an agent, which dynamically generates and executes XR interactions in response to user needs, with minimal predetermined logic.

\subsection{Example Application - Fitness Instruction and Feedback}
Exercising without professional oversight increases the risk of improper technique and injury. XR-AI systems can analyze sensing data against expert guidelines to enable safer, more effective workouts~\cite{mandic2023arfit, ma2024avattar, liao2025golf}. This example shows an AI assistant delivering instructions and personalized feedback for bicep curls (\cref{fig:dumbbell}). Users initiate and stop tracking with a ``V'' gesture. An XR agent evaluates tracked movements and generates spatial visualizations and speech feedback.

\subsubsection{Developer-Defined Interaction} 
Users employ a V gesture to start tracking. The same gesture stops tracking and triggers AI analysis. When hand motion tracking is active, the system displays a virtual dumbbell in the user's right hand, as shown in~\cref{fig:dumbbell}(a). In this example, developers retain control over interaction and data collection.

\subsubsection{Agent-Generated Feedback}
The agent analyzes hand position and rotation data, identifies the most critical error using heuristics retrieved from a web search, and then generates corrective feedback through a combination of text labels, geometric primitives, and speech, as shown in~\cref{fig:dumbbell}(b). The layout, elements, and content of the feedback composition fully depended on user performance and agent decisions at runtime.

\section{Using the XARP MCP Server}
\label{sec:how_to_mcp}

XARP supports no-code XR-AI prototyping workflows, allowing developers and end-users to orchestrate agentic tools via natural language and MCP~\cite{mcp}. XARP exposes previously unavailable runtime XR capabilities to the broader MCP ecosystem of third-party agents and tool providers, enabling agents to sense and act directly on an XR device. The how-to scenario demonstrates how to launch the XARP built-in MCP server and connect it to a local large language model (LLM) chat interface alongside other MCP servers. Inspired by recent work on generative AI for XR authoring~\cite{giunchi2024dreamcodevr, de2024llmr, zhu2025agentar}, we present a minimal example showcasing XR app modification via natural language interaction with an AI agent.

\subsection{How-To Scenario - XR MCP Workflow}

\subsubsection{Starting the Built-in XR MCP Server}

After installing XARP as described in~\cref{sec:how_to}, the user launches the built-in MCP server using \texttt{xarp\_mcp <name> <port>}. This command creates a XARP server and associates it with an MCP server awaiting external client connections. Calls to the MCP server are handled as XARP procedures by the XR device.

\subsubsection{Connecting an Agent to the Built-in XR MCP Server}

\begin{figure}[t]
\centering
\includegraphics[width=\linewidth]{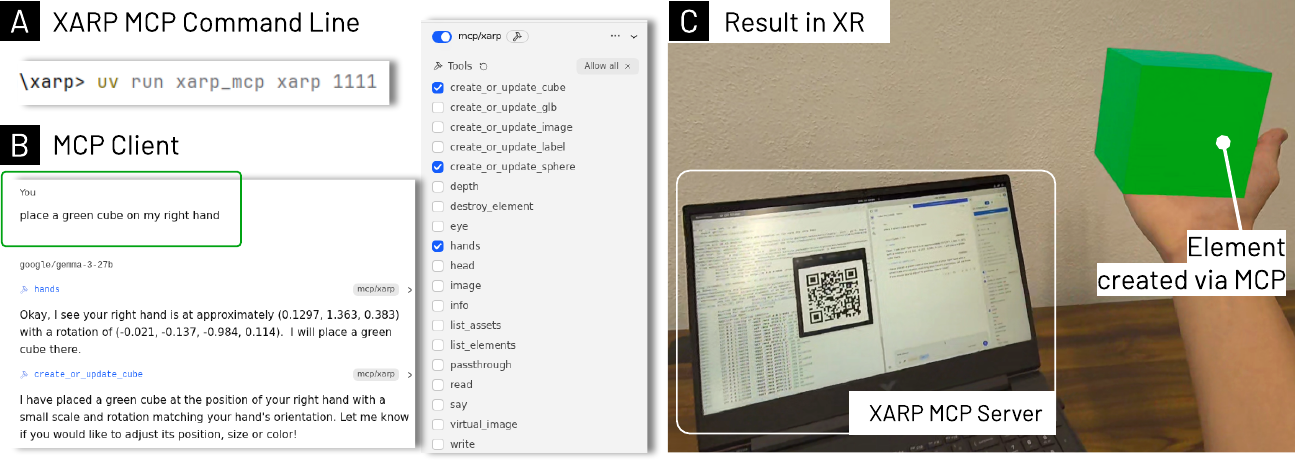}
\caption{XARP built-in MCP server. (a) The XARP built-in MCP server is launched via command line, forwarding incoming tool calls to a connected XR device. (b) An MCP client connects to the server using the server address, allowing agents to invoke XARP tools as needed; users interact in natural language through chat interfaces, enabling live no-code XR prototyping directly on the XR device. (c) Results are visualized live on the connected XR device.}
\label{fig:mcp}
\Description{Figure illustrating the built-in XARP MCP server for no-code XR prototyping. Panel (A) shows a command-line interface launching the XARP MCP server with a terminal command. Beside it is a tool list interface displaying available MCP tools such as create or update sphere, image, label, and cube elements. Panel (B) shows a chat-based MCP client interface where a user requests ``place a green cube on my right hand,'' and the assistant responds by describing the detected hand position and confirming creation of a green cube attached to the user’s hand orientation. Panel (C) shows the result in XR: a user holds up their hand with a floating green cube anchored above it, while a nearby laptop displays the XARP MCP server and a QR code connection interface. The figure demonstrates how natural-language MCP interactions can create and manipulate XR elements live on a connected XR device.}
\end{figure}

In this how-to, we use LMStudio\footnote{LMStudio: \url{https://lmstudio.ai}. Accessed March 29, 2026.}, a local LLM runtime with a chat interface. After starting a new chat with a loaded model, MCP servers can be added to augment the model's capabilities. In LMStudio, this requires editing a JSON configuration file with the XARP MCP server URL. After that, the model can invoke XARP functions as needed. \cref{fig:mcp} shows a simple example of a user interacting via an MCP client chat interface with a model capable of executing XR operations through the built-in XARP MCP server.

\section{Developer API}
\label{sec:api}

\begin{figure}
\centering
\includegraphics[width=1\linewidth]{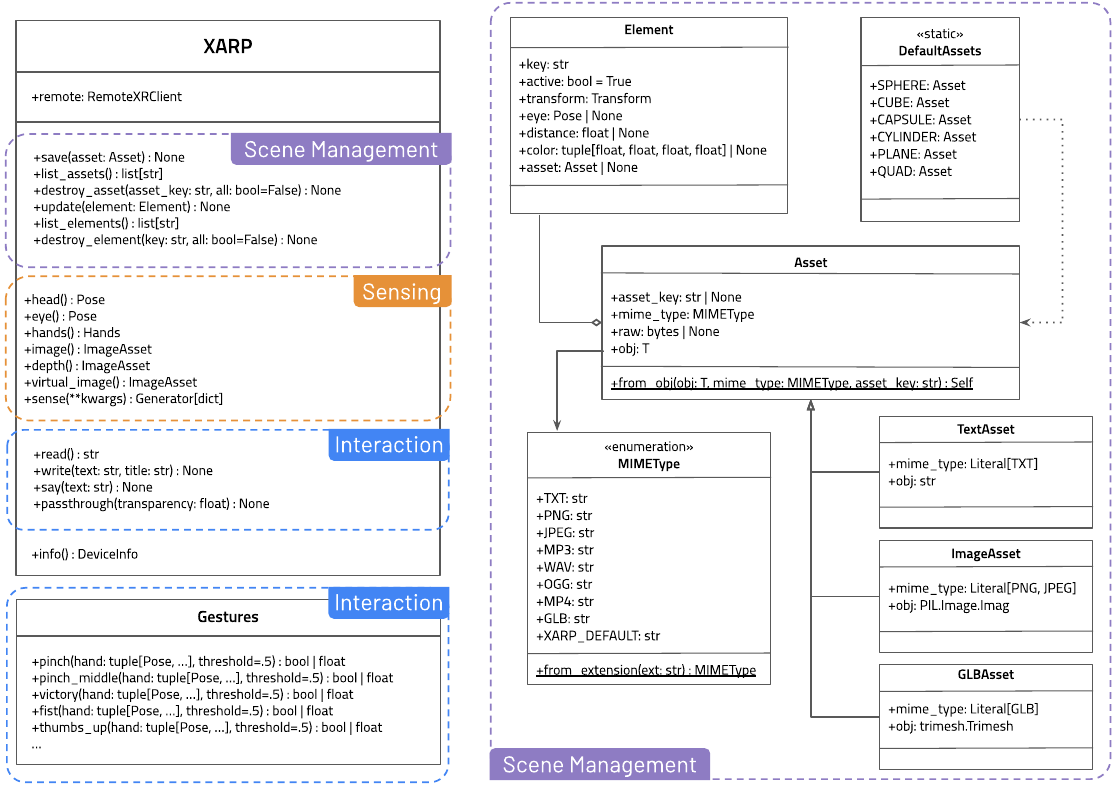}
\caption{UML class diagram of the XARP developer Python API.}
\label{fig:python_api}
\Description{UML class diagram of the XARP Python API. The diagram is divided into functional regions for scene management, sensing, and interaction. On the left, the XARP class provides methods for scene management such as save, list assets, destroy asset, update element, list elements, and destroy element. It also provides sensing methods including head pose, eye pose, hand tracking, RGB image capture, depth sensing, virtual image capture, and a generic sensor stream interface. Interaction methods include reading text input, writing text output, speech synthesis, and passthrough transparency control. A separate Gestures module defines gesture-recognition functions such as pinch, pinch hold, victory gesture, fist, and thumbs up. On the right, the scene management data model defines an Element class with properties including transform, visibility, type, distance, color, and asset references. Elements are associated with Asset objects that store MIME type, raw bytes, and object data. Asset subclasses include TextAsset, ImageAsset, and GLBAsset for text, image, and 3D mesh content. A MIMEType enumeration lists supported formats such as TXT, PNG, JPEG, MP3, WAV, OGG, MP4, GLB, and a default XARP type. A static DefaultAssets class provides built-in primitive assets including sphere, cube, capsule, cylinder, plane, and quad geometries. The figure illustrates the modular structure of the XARP Python API and relationships between XR scene elements, sensing interfaces, interaction utilities, and asset representations.}
\end{figure}

Developers interact with XARP primarily through a \emph{facade}~\cite{gamma1995designpatterns} that abstracts data validation, serialization, transmission, and deserialization, while providing convenient default values. Both synchronous and asynchronous (\texttt{asyncio}) facade variants are available and share the same underlying mechanisms. Developers are not limited to the facade layer and may access lower-level abstractions directly, down to the WebSocket transport.~\Cref{fig:python_api} shows a UML class diagram of the XARP Python API.

\subsection{Scene Management}
XARP's scene management system is based on two main concepts: \texttt{Elements} and \texttt{Assets}.

\subsubsection{Elements}
Elements are state change \emph{commands}~\cite{gamma1995designpatterns} that modify unique virtual entities on the client device. Applications transmit only property changes rather than full states, keeping network messages compact and avoiding retransmission of large assets. As elements are modified server-side, changes do not take effect until \texttt{xarp.update(element)} is called. This method has \emph{upsert} semantics: if an Element with the provided key exists, it is updated; otherwise, a new one is created. \texttt{xarp.list\_elements} lists all existing Element keys, and \texttt{xarp.destroy\_element()} destroys a specific Element or all Elements. On the client side, Elements are implemented as Unity game objects, making position, rotation, scale, and parenting transforms inherently available.

\subsubsection{Assets}
Assets are objects representing reusable content that can be rendered to the user. Currently, XARP supports UTF-8 text, JPG and PNG images, OGG audio, GLB 3D models, MP4 video, and 3D primitives. Each asset type has a MIME type that the client uses to decide how to render it using a \emph{state}~\cite{gamma1995designpatterns} pattern on the Element. Assets hold bytes of their raw content, but can also be manipulated in Python through their \texttt{obj} property, which exposes convenient object types for each asset type, such as Pillow\footnote{Pillow: \url{https://pillow.readthedocs.io/en/stable}. Accessed March 27, 2026.} images or Trimesh\footnote{Trimesh: \url{https://trimesh.org}. Accessed March 27, 2026.} meshes. 

Assets can be reused across Elements and stored persistently on the client using \texttt{xarp.save\_asset}, so applications transmit an asset only once and reuse its \texttt{asset\_key} to associate it with multiple Elements. All stored assets can be listed using \texttt{xarp.list\_assets}, and individual assets can be deleted by \texttt{asset\_key} or all at once using \texttt{xarp.destroy\_asset}.

\subsection{Sensing}
Applications using XARP have easy access to the wide range of sensors available on XR devices.

\subsubsection{Head and Eye}
\texttt{xarp.head} and \texttt{xarp.eye} both return a \texttt{Pose} object containing position and orientation. \texttt{xarp.head} tracks the user's device, whether handheld or head-mounted, while \texttt{xarp.eye} tracks the RGB camera, matching the user's point of view or, for head-mounted devices, the left eye camera. These poses can be used to track user movement over time or place Elements relative to the user.

\subsubsection{Hands}
\texttt{xarp.hands} returns a \texttt{Hands} object with \texttt{left} and \texttt{right} attributes, each holding a tuple of joint positions and orientations. If a hand is not tracked, the respective tuple is empty.

\subsubsection{RGB, Depth, and Virtual Images}
XARP provides three image capture methods, each returning an \texttt{ImageAsset}: \texttt{xarp.image} for RGB, \texttt{xarp.depth} for depth, and \texttt{xarp.virtual\_image} for the rendered virtual scene. Capture extrinsics vary by device; intrinsics are available via \texttt{xarp.info} (see~\cref{sec:device-info}).

\subsubsection{Sense Stream}
All sensing methods can be called individually to obtain the latest sensor reading. For high-frequency repeated readings, \texttt{xarp.sense} is recommended instead, as it requests the client to stream sensor data continuously until stopped, reducing inter-reading delay. The method accepts boolean keyword flags, one per sensing method, selecting which sensors to include in each reading, and returns a Python generator for iterating over successive readings. Once the stream is no longer needed, the application should call \texttt{close} on the generator to notify the client to stop transmitting.

The \texttt{rt} (real-time) flag controls reading behavior. With \texttt{rt=False}, every frame is yielded in sequence with minimal inter-frame delay, though readings may lag behind the current state if server-side processing is slow; this mode is suited for continuous data collection. With \texttt{rt=True}, only the latest frame is yielded and older frames are dropped, keeping memory usage minimal and readings current at the cost of potential discontinuities; this mode is preferred for interactive applications.

\subsection{Interaction}

\paragraph{Read, Write \& Say}
The method \texttt{xarp.read} acquires text input from the user, either through keyboard entry or dictation, depending on the capabilities of the client device. This method blocks the caller until the client transmits the input. The method \texttt{xarp.write} displays text to the user in a panel and is suitable for presenting notifications. The method \texttt{xarp.say} accepts a string argument, synthesizes speech from it, and plays the resulting audio on the client device. This method blocks the caller until audio playback is complete.

\paragraph{Gestures}
XARP supports a set of static hand gestures detected from hand sensor readings. Supported gestures include: pinch with thumb and any finger, thumbs up, index--middle ``V'', index--thumb ``L'', closed fist, open palm, and pointing. Each gesture function computes an internal metric that characterizes the gesture; for example, \texttt{xarp.pinch} measures the distance between the thumb tip and index tip and returns \texttt{true} if this distance falls below an empirically defined threshold, and \texttt{false} otherwise. To retrieve the raw metric value instead of the boolean result, the caller passes \texttt{threshold=None} as an argument.

\paragraph{Passthrough}
The method \texttt{xarp.passthrough} accepts a value in the range from zero to one and controls the opacity of the virtual environment background. A value of zero renders the background fully transparent, allowing the user to see the physical environment through the XR device. A value of one renders the background fully opaque, replacing the physical environment with a solid color background.

\subsection{Device Info}
\label{sec:device-info}
While XARP abstracts device-specific details to enable multiplatform development, it also provides access to device-specific information through the method \texttt{xarp.info}, which returns a data structure containing information such as the device model and intrinsic camera parameters. Because device capabilities vary and may not fully cover the XARP abstraction, this method can also be used to query which XARP methods the client supports, allowing applications to handle unsupported features gracefully and define fallbacks for broader compatibility.

\subsection{Spatial Data Types}
XARP provides a set of common data types for spatial operations. \texttt{Vector3} represents a point or vector in three-dimensional space. \texttt{Quaternion} represents an orientation. Both types support standard algebraic operations. \texttt{Pose} composes a position of type \texttt{Vector3} and a rotation of type \texttt{Quaternion}; a \texttt{Pose} can define a ray and compute the position of a point at a given distance along it. \texttt{Transform} extends \texttt{Pose} with an additional scale component of type \texttt{Vector3}, representing the scale of an object along each axis.

\section{Architecture}
\label{sec:arch}

\begin{figure}
    \centering
    \includegraphics[width=0.9\linewidth]{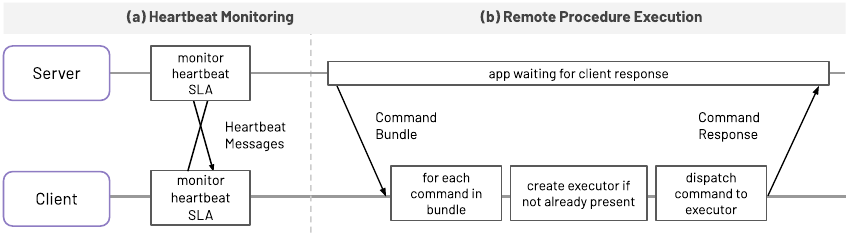}
    \caption{XARP client-server interaction. (a) Heartbeat Monitoring: XARP uses a client-server architecture with session management functionalities to ensure that the session ends gracefully in the event of a disconnection, and command bundles are issued by the server to ensure execution in the same frame on the client. (b) Commands are sent from the server in bundles, potentially containing multiple commands. The client dispatches each of these commands to an appropriate executor that handles the actual execution on the XR device.}
    \label{fig:architecture}
    \Description{Diagram illustrating the XARP client-server interaction model. The figure is divided into two parts: (a) heartbeat monitoring and (b) remote procedure execution. In part (a), a server and client each monitor heartbeat service-level agreements through exchanged heartbeat messages to detect disconnections and manage session integrity. In part (b), the server sends command bundles to the client while the application waits for client responses. On the client side, each command in the bundle is processed sequentially: the client checks each command, creates an executor if one does not already exist, dispatches the command to the executor, and returns a command response to the server. The diagram illustrates how XARP coordinates synchronized remote execution and fault-tolerant communication between a server application and an XR device client.}
\end{figure}

This section presents the internal workings of XARP and contains relevant information for researchers and engineers aiming to modify or replicate the toolkit. Developers do not need to interact with the concepts in this section since they operate through the high-level API described in~\cref{sec:api}. XARP implements a client-server architecture that shifts the application logic from XR devices, which is currently the mainstream approach for XR development, into a Python server. The Python application uses XARP to issue remote procedure calls and execute XR functions in a standardized client application deployed to the XR device.~\Cref{fig:architecture} illustrates this client-server interaction.

\subsection{Server}

\subsubsection{Session Management}
The XARP server is built on FastAPI with asynchronous programming, suited for the I/O-bound workload of maintaining a WebSocket connection with the remote XR device. When a client connects, XARP instantiates a session object. The session manages the application lifecycle, coordinating both inbound and outbound communication over the connection.

The session runs a dedicated async task that monitors a configurable heartbeat SLA. If the client is silent beyond this threshold, the task triggers a graceful shutdown, preventing both sides from waiting indefinitely and releasing resources held by the unresponsive client. Any message exchanged during normal application flow also resets the SLA timer.

The session also runs a dedicated async task that periodically sends heartbeat messages, keeping the connection alive when the server is silent. A third async task reads from the WebSocket connection and fans out incoming messages to the rest of the system, fulfilling pending results and resetting the heartbeat SLA. Once initialized, the session is wrapped by a facade object, described in~\cref{sec:api}, which is then passed to the application.

\subsubsection{Command Bundles}

Commands are grouped into bundles, the atomic unit of transmission between server and client. All commands in a bundle are guaranteed to execute within the same client frame, minimizing latency between them. Bundles are serialized in MessagePack, a binary format more efficient than JSON. The developer facade does not currently expose bundle creation directly; however, developers who need to guarantee minimal delay between two commands can bypass the facade and create bundles manually through the session.

\subsubsection{Response Modes}

\begin{figure}
    \centering
    \includegraphics[width=0.85\linewidth]{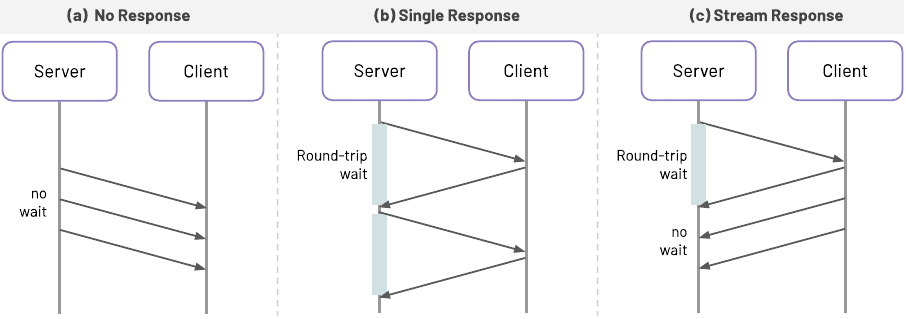}
    \caption{Command bundles from the server may either (a) resolve instantly after being sent, (b) receive a single response asynchronously, or (c) receive a stream of responses to a single command.}
    \label{fig:responses}
    \Description{Sequence diagrams illustrating three XARP command-response interaction patterns between a server and client. Panel (a), labeled ``No Response,'' shows the server sending commands to the client without waiting for any reply, allowing immediate continuation of execution. Panel (b), labeled ``Single Response,'' shows the server sending a command and waiting for a single asynchronous response from the client before continuing, introducing a round-trip wait period. Panel (c), labeled ``Stream Response,'' shows the server sending a command that produces a continuous stream of responses from the client; after the initial round-trip wait, additional streamed responses are delivered without further blocking. The figure demonstrates how XARP supports asynchronous command execution with optional one-time or streaming responses between the server and XR client.}
\end{figure}

Bundles support three response modes, illustrated in~\cref{fig:responses}: no-response, single-response, and stream. No-response bundles resolve immediately after being sent, requiring no acknowledgment from the client. They are suited for fire-and-forget actions such as \texttt{xarp.write} and \texttt{xarp.passthrough}, where the server does not need to await their effect. Single-response bundles create a Python future that the application awaits. When the client finishes executing the bundle, it sends a response back to the server. The session receives it in a dedicated async task, matches it to the waiting caller via a correlation identifier, and fulfills the future. Stream bundles yield results continuously through a Python generator that the application iterates. Upon receiving a stream bundle, the client maintains an internal queue and returns at most one result per frame. Either side can terminate a stream: the client sends an EOS (End-of-Stream) message, while the server closes the generator, which triggers a cancellation message to the client.

\subsection{Client}

The XARP client is a Unity XR application that mirrors the server-side session. It runs asynchronous tasks that monitor the server heartbeat SLA, send periodic heartbeats, receive bundles from the server, and send back responses.

\subsubsection{Command Handling}

Once a bundle is received, the client session dispatches its commands to Executors, Unity GameObjects responsible for executing commands with the given parameters. Each server command maps to a corresponding Executor. For example, \texttt{xarp.read} is handled by \texttt{ReadExecutor}, which manages the platform-specific details of displaying a keyboard and returning a response once the user finishes typing. Executors are instantiated lazily on first use and reused for subsequent calls to the same command. Command-to-Executor mappings are configured through a ScriptableObject and can be remapped in the Unity Editor without code changes. Executors that depend on platform-specific features are isolated in separate C\# namespaces to simplify porting. XARP currently ships a Meta Quest 3 reference implementation and a partial Android ARCore implementation. Full clients for additional platforms are on the roadmap.

\subsubsection{Connection Management}

Clients connect to an XARP server by scanning a QR code that encodes the server address and named parameters accessible to the application. On disconnection, the client attempts to reconnect to the last known address using a backoff-jitter strategy to avoid overloading the server. This auto-reconnection behavior enables the live reloading feature described in~\cref{sec:how-main-workflow}. Scanning a new QR code connects to a different server, allowing users to switch between XARP applications without typing addresses.

\section{Related Work}
\label{sec:rw}
Our goal is to facilitate prototyping of XR-AI systems for developers and researchers who are fluent in Python, the dominant language of contemporary AI research~\cite{maslej2025aiindex}, but unfamiliar with XR development stacks based on C\#/C++ and game engines. Achieving this requires lowering the entry barriers imposed by conventional XR development workflows, while preserving the ability to modify and integrate XR interfaces with experimental systems. Prior work has addressed these requirements in isolation, but reconciling accessibility with programmatic control in a single toolkit remains an open challenge. Moreover, Python tools for XR development remain scarce, forcing developers and researchers to bridge ecosystems. XARP addresses this gap with a Python library of common XR operations, simplifying the development and deployment of XR-AI prototypes and offering a high degree of control. These functionalities are exposed as callable tools and MCP servers for integration with AI agents.

\subsection{XR Authoring Tools}

Prior work has proposed XR authoring tools that make XR content creation accessible to users unfamiliar with traditional XR development, utilizing domain-specific low- or no-code abstractions to simplify content creation. However, these abstractions either remove programmatic control entirely or still require conventional XR workflows for integration with customized research systems.

DART~\cite{macintyre2004dart} is a pioneering AR authoring toolkit that allows designers to rapidly turn storyboards into working AR experiences by specifying relationships between physical and virtual elements through the capture and replay of synchronized video and sensor data.
Teachable~Reality~\cite{monteiro2023teachable} is a web-based mobile AR prototyping tool that enables users to create functional tangible AR interactions with everyday objects by demonstrating them to an on-demand vision classifier and linking recognized states to AR outputs through a trigger-action interface, without programming or predefined markers.
Mucho~\cite{leiva2025mucho} is a no-code immersive prototyping tool for multimodal XR interaction that supports design-by-enacting, letting designers demonstrate inputs such as hand gestures, proximity, speech, and gaze into a timeline, add system actions during playback, and automatically generate a runtime state machine for testing the resulting interactive experience.
ProInterAR~\cite{ye2024prointerar} is an integrated visual programming platform that enables novice AR creators to build general-purpose immersive AR applications by authoring real and virtual content through an AR-HMD and scripting their interactive behaviors with Scratch-inspired blocks on a tablet, without text-based coding.

\subsection{Live XR Authoring with AI Agents}
Recent research has proposed live XR authoring systems that allow users to describe XR experiences in natural language to an LLM-based agent that generates and executes the corresponding code on the fly, offering an intent-based workflow similar to the one demonstrated in \cref{sec:how_to_mcp}. This approach benefits from the accessibility and expressiveness of natural language as an authoring abstraction, but does not afford direct edits to the generated artifacts. Researchers and developers integrating XR into experimental systems must export generated artifacts to a conventional XR development tool for fine-grained editing.

DreamCodeVR~\cite{giunchi2024dreamcodevr} captures spoken behavior descriptions, generates corresponding C\# Unity scripts, and executes them live in a running VR application.
LLMR~\cite{de2024llmr} coordinates a pipeline of specialized agents for planning, scene analysis, skill grounding, code generation, and self-inspection to let users construct and modify interactive mixed reality scenes with natural language prompts at runtime.
Ostaad~\cite{aghel2024people} is a conversational agent that allows non-programmers to design interactive VR scenes through embodied prompts.
AgentAR~\cite{zhu2025agentar} is a tool-augmented LLM agent that supports in-situ authoring of end-to-end interactive AR applications from natural language inputs, and can produce systems across several application categories.
LLMER~\cite{chen2025llmer} follows a similar pattern, but instead of code, it generates structured JSON that is dispatched to preconstructed runtime modules.
VRCopilot~\cite{zhang2024vrcopilot} and ImaginateAR~\cite{lee2025imaginatear} enable immersive authoring through multimodal natural-language interaction with generative AI, targeting indoor furniture layout co-creation and on-site outdoor 3D asset generation respectively, with less emphasis on interactivity.

\subsection{XR Toolkits with Game Engine Workflows}
Prior research has proposed toolkits built on top of game-engine workflows that reduce the accidental complexity of XR development~\cite{brooks1987silver} without sacrificing developer control, typically by introducing abstractions that simplify laborious steps or accelerate the construction of specific classes of XR systems. These toolkits, however, still require familiarity with conventional game-engine workflows, effectively decoupling AI researchers from their typical stack.

Ubiq~\cite{friston2021ubiq} is an open-source Unity toolkit for social MR, providing room management, voice, avatars, and component-based messaging. Ubiq-Exp~\cite{steed2022ubiq} builds on this by providing tools for remote and distributed MR studies. It integrates specialized features such as distributed logging, record-and-replay, and in-world questionnaires.
AUIT~\cite{evangelista2022auit} is a Unity extension that lets developers compose adaptive XR UIs from declarative objectives such as visibility and reachability, with a multi-objective solver resolving conflicts between them in real time.
XRtic~\cite{muthukumarana2022xrtic} is a hardware-software prototyping toolkit comprising modular SMA cloth actuators, a controller bus system, and a Unity C\# scripting interface for triggering clothing deformations synchronized with virtual content events.
XRSpotlight~\cite{frau2023xrspotlight} is a Unity Editor panel that automatically abstracts existing XR scene implementations into natural-language trigger-action rules, helping novice developers inspect interaction logic, identify similar examples, and transfer interaction behavior across scenes and toolkits through copy-paste.
ConnectVR~\cite{chen2024connectvr} is a no-code trigger--action authoring Unity plugin that enables non-technical artists and storytellers to create agent-based VR narratives by connecting player actions to virtual character and object behaviors, adding interactivity to scenes assembled in the Unity Editor.
CoCreatAR~\cite{numan2025cocreatar} is an asymmetric collaborative AR authoring system that connects an in-situ mobile app for on-site testing, spatial capture, and contextual annotation with a Unity-based ex-situ editor, allowing remote developers to refine site-specific outdoor AR experiences using real-time feedback from the target environment.
MCP-Unity~\cite{wu2025mcp} connects an LLM-based agent to the Unity Editor via MCP for project authoring, which can later be built and deployed to the target XR device.

\subsection{Emerging XR Ecosystems}

Recent research has proposed XR toolkits targeting emerging XR ecosystems, offering alternatives beyond the mainstream C\#/C++ game-engine stack. Despite their flexibility, these alternatives lack robust integration with the Python ecosystem and AI agents.

WebXR\footnote{WebXR: \url{https://www.w3.org/TR/webxr/}. Accessed May 10, 2026.} is a W3C browser API that gives JavaScript applications direct access to XR hardware, including headset poses, motion controllers, and stereo rendering, without a native runtime or game engine. p5.xr~\cite{p5xr} extends p5.js~\cite{p5js}, a JavaScript creative coding library, to run sketches in AR and VR via WebXR, lowering the entry barrier for developers familiar with JavaScript but not 3D development. XR Blocks~\cite{li2025xrblocks} introduces a ``Reality Model'' that abstracts users, environments, and agents into first-class primitives, leveraging native AI execution (via TensorFlow and Gemini) to unify spatial perception and interaction. In contrast, IWSDK~\cite{iwsdk} employs a high-performance ECS architecture integrated with an MCP-connected AI agent, facilitating autonomous, closed-loop development through real-time code generation, scene graph inspection, and self-debugging. While these WebXR toolkits bypass the traditional build-deploy cycle, they remain isolated from the Python ecosystem central to contemporary AI research.

NVIDIA CloudXR~\cite{cloudxr} is a GPU-accelerated streaming platform that decouples XR rendering from display by running OpenXR applications on high-performance servers and streaming encoded video to lightweight client devices over standard networks. CloudXR Runtime acts as an OpenXR-compliant runtime on Windows and Linux, capturing stereo frames, encoding them via hardware-accelerated codecs, and transmitting them alongside audio and tracking data to native clients on several XR devices and web browsers via CloudXR.js. Server applications are built against the OpenXR C API or integrated through game engines. CloudXR does not target Python developers and does not expose XR functionality to AI agents.

Godot\footnote{Godot Engine: \url{https://godotengine.org/}. Accessed May 10, 2026.} is an open-source game engine that offers XR support through Godot~XR~Tools~\cite{godotxrtools}, a library of reusable XR interaction components including locomotion, object pickup, pointer, and snap zones. Two experimental plugins, py4godot~\cite{py4godot} and godot-python-extension~\cite{godotpythonextension}, add Python as a scripting language inside the editor. This combination, however, does not guarantee portability with the Python ecosystem, as packages with binary dependencies may be incompatible with standalone XR hardware. Development depends on the build-deploy cycle of conventional game engine workflows, and XR functionality is not natively exposed to AI agents.

PyOpenXR~\cite{pyopenxr} provides Python bindings for the OpenXR API and supports tethered VR prototyping on Windows and Linux. It closely exposes the OpenXR C API in Python, allowing developers to manage extension negotiation and integrate custom render loops. This level of control is useful, but it requires platform-specific expertise that is typically outside the scope of AI research workflows. Its reliance on a PC-based runtime also limits standalone XR deployment, and it does not provide built-in support for AI agent integration.

SnapNCode~\cite{wei2025snapncode} takes a different approach. Rather than targeting the XR runtime layer, it is a browser-based spatial IDE that lets programmers capture physical object states from a live video stream and embed them directly into Python code as images, with object-state changes triggering attached code snippets opportunistically via mobile or headset cameras. SnapNCode primarily contributes a reduced ``perceptual distance'' between physical state and source code. However, its spatial functionality is limited to 2D video input and object-proximity operators such as \texttt{In()} and \texttt{On()}, which are applied within 2D bounding boxes; consequently, it lacks support for true 3D spatial reasoning or immersive hardware rendering.

\section{Formative Case Studies}
\label{sec:formative}

To gain first-hand insight into novice XR developers' experience and directly observe pain points faced by this audience, we conducted two case studies with high-school seniors as part of a summer research program at our home university. They built XR-AI systems using a mix of conventional XR workflows and an early version of XARP.

\subsection{AI Fencing Coach}

This case study produced an AI agent that provides sabre fencing feedback on off-the-line movements based on biomechanical features elicited with fencing coaches~\cite{kessler2025sabre}. The system depended on full-body tracking beyond what the HMD available for this project offered. To overcome this, the student considered integrating a mounted smartphone for tracking full-body motion in a third-person perspective. The complexity of integrating that external device and synchronizing the incoming data was incompatible with the project's time frame, and they pivoted to analysis of a public dataset of prerecorded bouts.

\paragraph{Impact on the toolkit design} This case study shaped two design choices in XARP, both aligned with~\citet{raffaillac2022researchers}'s recommendations for research-oriented toolkits.
Following the \textit{duplicate} principle, which recommends against assuming that elements like XR devices are singular, XARP allows multiple devices to connect to the same server and easily exchange data through simple shared variables.
Following the \textit{accumulate} principle, which calls for retaining interaction data in interoperable formats, XARP data types extended Pydantic models~\footnote{Pydantic Models: \url{https://docs.pydantic.dev/latest/concepts/models/}. Accessed March 27, 2026.} for easy data validation and serialization.

\subsection{Drone Assistant}

This case study produced an AI assistant for outdoor pick-and-place embodied as a virtual drone~\cite{narchetty2025drone}. The system required integration with drone flight simulators, AI agents, and, in future work, hardware platforms. A key challenge was synchronizing the drone's virtual avatar, simulation states, and AI agent perception-action cycle.

\paragraph{Impact on the toolkit design} This case study motivated two design choices. 
First, we strive to keep XARP flexible without enforcing a particular system integration architecture. This allows developers to leverage the rich Python ecosystem to integrate with external services in any form, including third-party libraries and SDKs, bindings for robotics and simulation platforms, and AI integration beyond API calls through direct access to AI models, callable tools, and MCP.
Second, we allow state changes of persistent entities across layers (e.g., XR client, application logic, and simulations) using command objects that carry state deltas, implemented as Elements presented in~\cref{sec:api}.

\section{Early Acceptance Evaluation}
\label{sec:preliminary_eval}

After refining XARP based on insights from the formative process, we conducted an early acceptance evaluation to gather initial feedback from the target audience and assess the toolkit's acceptance potential. This evaluation consisted of a video walkthrough demonstration~\cite{ramakers2016retrofab, evangelista2021xrgonomics, ledo2018evaluation} followed by a UTAUT survey~\cite{venkatesh2003management} and a custom survey based on~\citet{ledo2018evaluation}. The videos covered both the fundamental concepts of XARP and practical aspects such as setup process and example implementations using key features of the toolkit, including tracking, speech interaction, AI agents for runtime generative behavior, and MCP integration. This method mirrors how developers often engage with new tools, by watching tutorials before deciding whether to try them in practice. This study was approved by our local institutional review board (UCSB \#39-25-0548)

\subsection{Participants}

We recruited 24 adults (ages 18–44; 14 male, 10 female, 0 non-binary) through social media and online AI-XR developer communities. All participants had at least one year of experience with AI, XR, or both. Participants were distributed across five countries, with varied educational and professional backgrounds. Most participants were trained in Computer Science, AI, or Engineering. XR experience varied from none (n=3) to more than five years (n=5). Participants most commonly reported experience with Meta Quest, iOS, and HoloLens. Unity with AR Foundation and the Mixed Reality Toolkit were frequently mentioned as development tools. In terms of AI experience, two participants reported no prior experience, while five reported more than five years of experience. Popular platforms included the OpenAI API, PyTorch, and TensorFlow.

\subsection{Measures}

\begin{figure}[t]
  \centering

  \begin{minipage}{\textwidth}
    \centering
    \hspace*{0.7cm}    \includegraphics[width=0.85\linewidth]{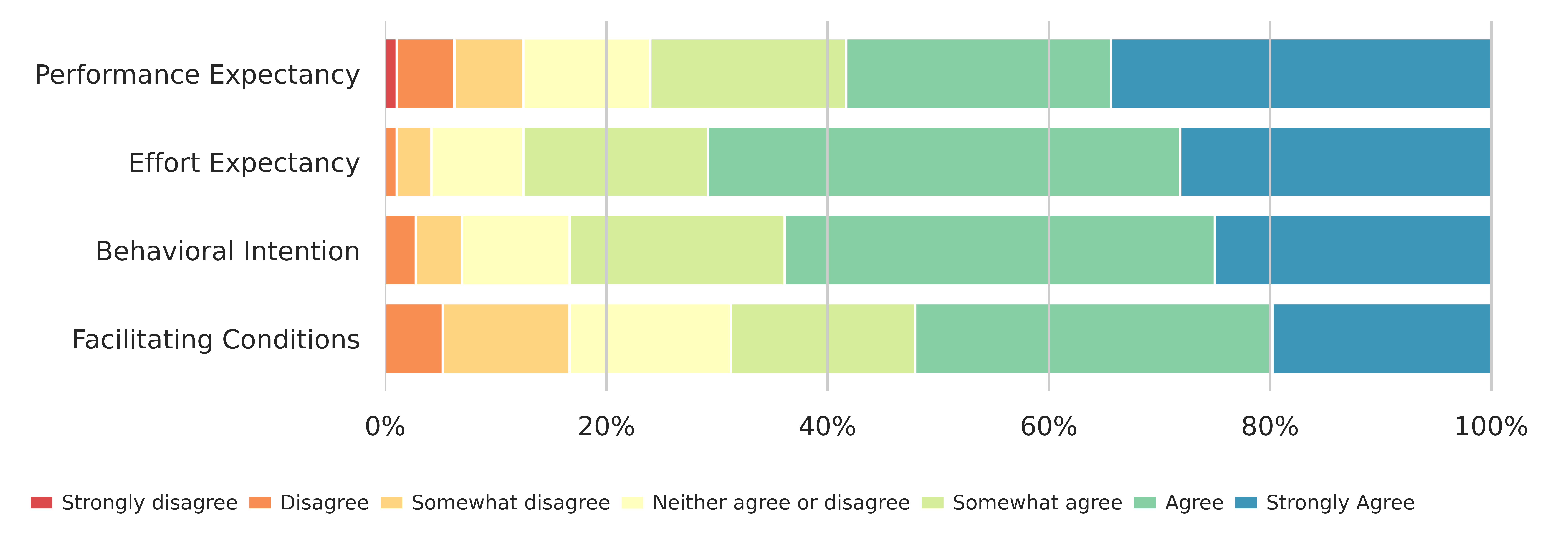}
    \caption*{(a)}
  \end{minipage}

  \vspace{1em}

  \begin{minipage}{\textwidth}
    \centering
    \includegraphics[width=0.9\linewidth]{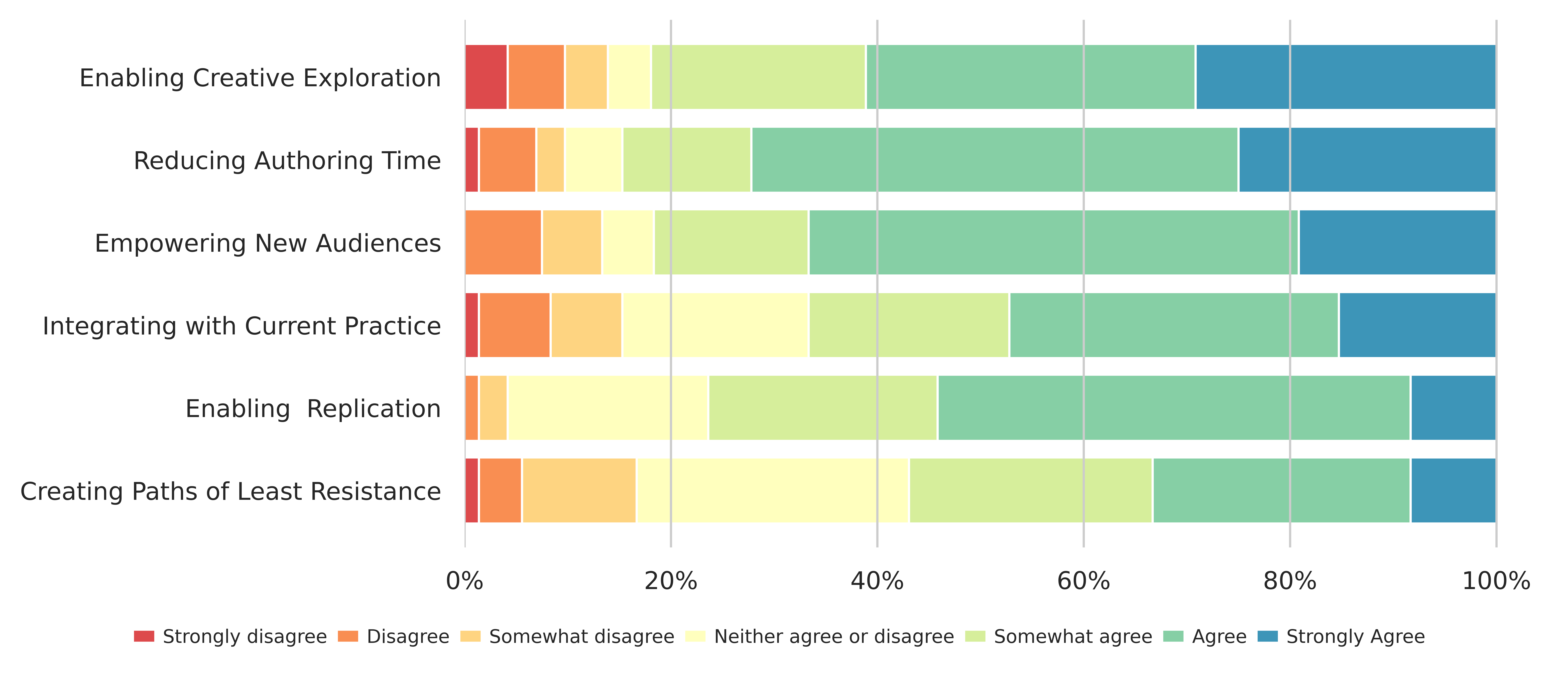}
    \caption*{(b)}
  \end{minipage}

  \caption{Distribution of item-level Likert responses to the four UTAUT constructs (a) and six HCITG constructs (b). Participant-level construct composites were significantly above the neutral midpoint.}
  \label{fig:likert}
  \Description{Figure showing stacked horizontal bar charts of participant survey responses for XARP usability and creative support constructs using Likert-scale ratings. Panel (a) summarizes responses for four UTAUT constructs: Performance Expectancy, Effort Expectancy, Behavioral Intention, and Facilitating Conditions. Each bar is divided into response categories ranging from strongly disagree to strongly agree. Most responses cluster in the agree and strongly agree categories, with relatively few disagreement responses. Panel (b) summarizes responses for six HCITG-related constructs: Enabling Creative Exploration, Reducing Authoring Time, Empowering New Audiences, Integrating with Current Practice, Enabling Replication, and Creating Paths of Least Resistance. These bars also show dominant agreement responses, especially for reducing authoring time, enabling replication, and integrating with current practice. Across both panels, neutral and disagreement responses occupy smaller portions of the distributions, indicating generally positive participant evaluations of XARP’s usability, accessibility, and creative workflow support.}
\end{figure}

We assessed XARP's \emph{potential acceptance} using the \textbf{Unified Theory of Acceptance and Use of Technology (UTAUT)} scale~\cite{venkatesh2003management} with 7-point Likert items. We excluded Social Influence because the study focused on individual early acceptance after a walkthrough rather than organizational adoption pressure. We complemented the survey with open-ended items asking why participants would adopt or avoid XARP.~\Cref{fig:likert} (a) shows the proportions of UTAUT responses.

To assess the \emph{perceived value} of the toolkit, we designed a custom survey based on the \textbf{HCI Toolkit Goals (HCITG)} framework compiled by Ledo et al.~\cite{ledo2018evaluation}, as no standardized instrument for this framework exists. We mapped each toolkit goal into a survey construct, further separating creative exploration from solution replication. Each HCITG survey construct was measured with three to six 7-point Likert-scale items, including at least one reverse-coded item. The survey covered the following constructs: \emph{Reducing Authoring Time}, \emph{Creating Paths of Least Resistance}, \emph{Empowering New Audiences}, \emph{Integrating with Current Practices and Infrastructure}, \emph{Enabling Replication of Existing Solutions}, and \emph{Enabling Creative Exploration}. Our customized HCITG survey is available in the supplementary material.~\Cref{fig:likert} (b) shows the proportions of HCITG responses.

We measured participant \emph{comprehension} of the walkthrough videos to ensure their judgments of the toolkit were based on informed understanding. Each video was followed by three to four comprehension-check items, along with a confidence check: ``The video provided enough evidence for me to answer the questions in this section.'' Our post-study questionnaire included the question ``What questions do you still have about the toolkit?'' to identify common comprehension gaps.

\subsection{Procedure}

The study was conducted fully online and asynchronously using Qualtrics\footnote{Qualtrics: \url{https://www.qualtrics.com/}. Accessed March 27, 2026.}, allowing participants to pause and resume at any point. The median completion time, including pauses, was 45 minutes, and the fastest participant finished the study in 25 minutes, consistent with our previous in-person pilot.

After providing informed consent and demographic information, participants were screened to ensure they were at least 18 years old and had at least one year of experience with AI or XR development. Eligible participants proceeded to watch seven short videos demonstrating XARP, presented in a fixed didactic order: \textit{Overview}, \textit{Setup}, \textit{Visual Q\&A}, \textit{Virtual Drone}, \textit{AI Agent}, \textit{MCP Integration}, and \textit{Modifying XARP}. Each video was approximately two minutes long, totaling around 15 minutes of content. Except for the overview video, which introduced core concepts and the toolkit architecture, each segment consisted of a brief introduction, a step-by-step walkthrough, and an example application. After each video, participants answered comprehension-check questions. Videos could be replayed while responding, as our goal was to ensure informed exposure to the toolkit and its capabilities, rather than assessing learning outcomes. Following the videos, participants completed the UTAUT and HCITG surveys. In this final section, they provided reasons for potentially adopting or avoiding XARP in their future AI-XR projects and could ask additional questions about the toolkit.

\subsection{Data Analysis}

All participants met the screening criteria for quality of responses: consistency on reverse-coded items, a minimum completion time of 20 minutes (15 minutes for video content plus 5 minutes for responses), open-text entries with more than a single word, and above-chance performance on comprehension checks.
We verified the reliability of items within the same construct with McDonald's $\omega$ before aggregating them into composite 0-1 scores.
We computed participant-level construct composites and used two-sided Wilcoxon signed-rank tests with Holm correction to assess deviations from the neutral midpoint.

Guided by the theoretical structure of UTAUT, we used multiple linear regression to examine associations between UTAUT constructs, moderators (age, gender, expertise), and \emph{Behavioral Intention}.
We analyzed participants' responses about reasons for potentially adopting or avoiding XARP, as well as their open-ended questions, using a rapid variant of the Framework Method~\cite{gale2013method}. This approach provides a transparent, systematic way to code and chart small corpora, enabling clear comparison across responses while minimizing analytic overhead.

\subsection{Results}

\subsubsection{Potential Acceptance}

\begin{figure}[t]
  \centering
  \begin{minipage}{0.44\textwidth}
    \centering
    \includegraphics[width=\textwidth]{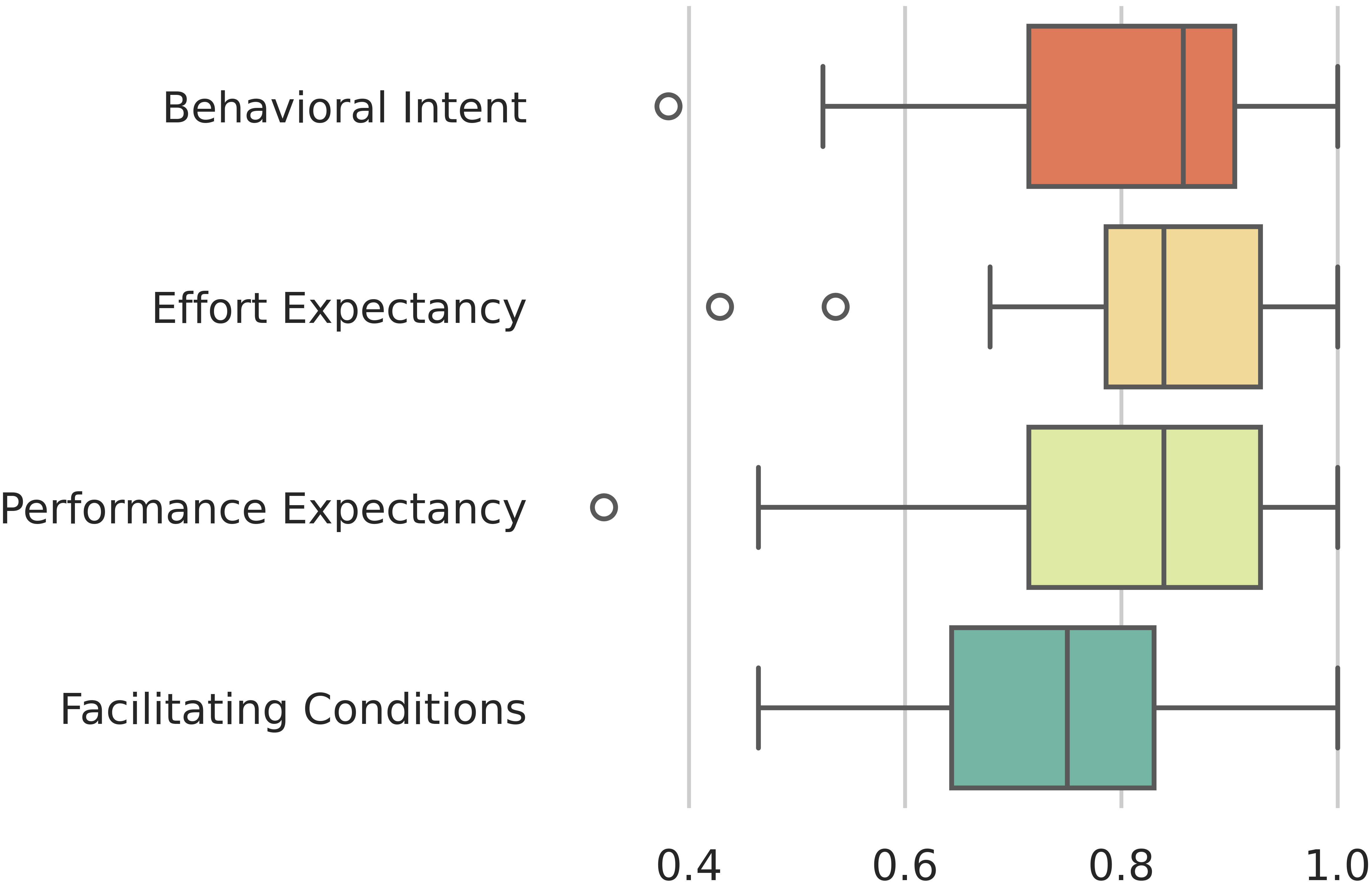}
    \caption*{(a)}
  \end{minipage}  \hfill
  \begin{minipage}{0.55\textwidth}
    \centering
    \includegraphics[width=0.9\textwidth]{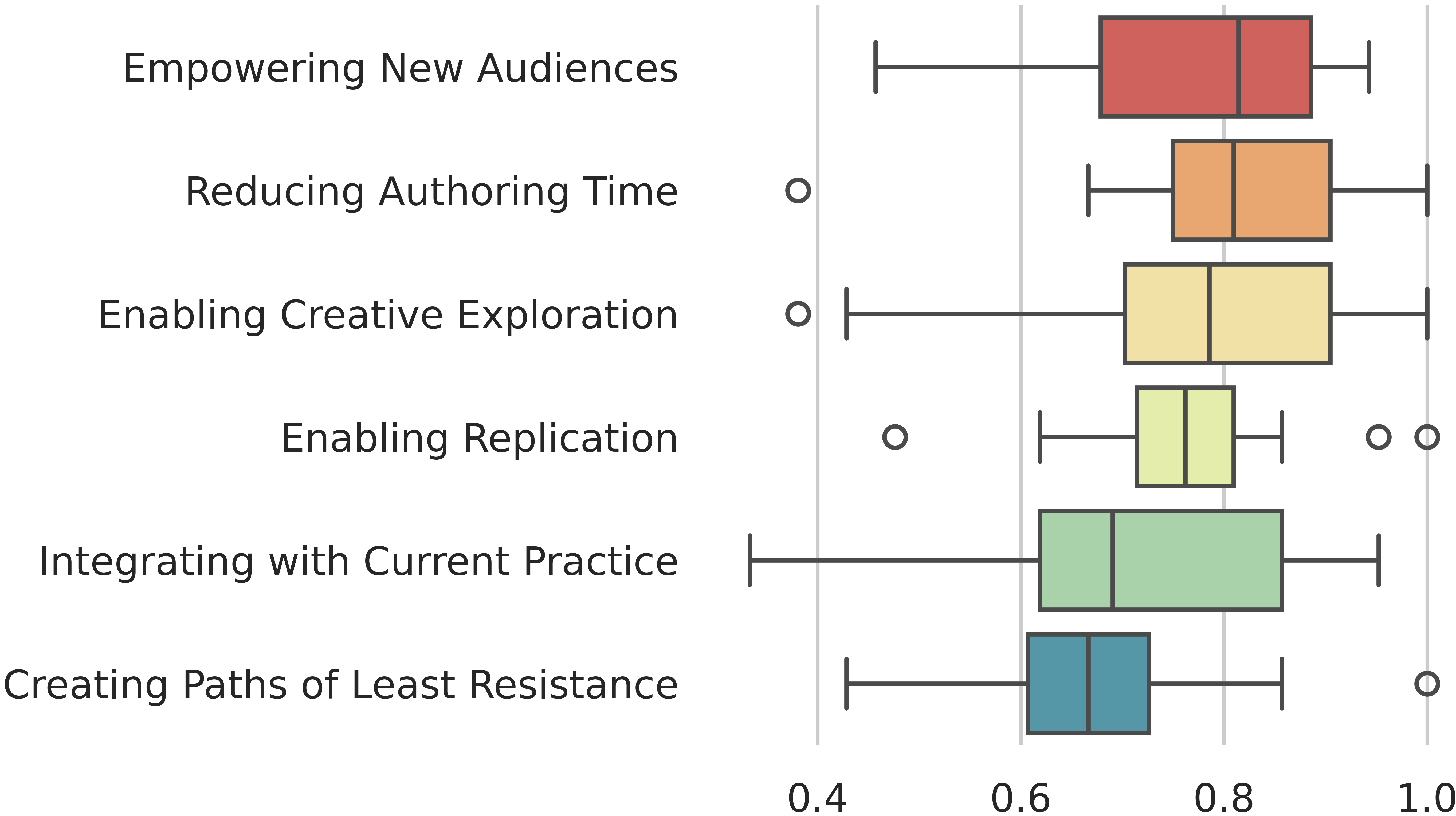}
    \caption*{(b)}
  \end{minipage}
  \caption{(a) UTAUT Score: \emph{Performance Expectancy} was a significant predictor of \emph{Behavioral Intention} on an exploratory multiple linear regression (b) Distribution of HCITG Scores.}
    \label{fig:boxes}
    \Description{Figure showing box plots summarizing participant construct scores for XARP evaluation metrics. Panel (a) presents box plots for four UTAUT constructs: Behavioral Intent, Effort Expectancy, Performance Expectancy, and Facilitating Conditions. Scores are shown on a normalized scale ranging approximately from 0.4 to 1.0. Median scores for all constructs are above 0.7, with Performance Expectancy and Effort Expectancy showing relatively high central tendencies. Several lower-score outliers are visible for Performance Expectancy and Effort Expectancy. The caption notes that Performance Expectancy significantly predicted Behavioral Intention in an exploratory multiple linear regression. Panel (b) presents box plots for six HCITG constructs: Empowering New Audiences, Reducing Authoring Time, Enabling Creative Exploration, Enabling Replication, Integrating with Current Practice, and Creating Paths of Least Resistance. Most distributions cluster between approximately 0.65 and 0.9, with Reducing Authoring Time and Empowering New Audiences exhibiting particularly high median scores. Enabling Replication shows a narrower distribution with a few high outliers near 1.0. Overall, the figure indicates consistently positive participant ratings across usability and human-centered innovation constructs.}
\end{figure}

Reliability of the UTAUT constructs was confirmed by McDonald's $\omega$: \emph{Performance Expectancy} ($\omega = 0.948,~4$ items); \emph{Effort Expectancy} ($\omega = 0.928,~4$ items); \emph{Facilitating Conditions} ($\omega = 0.928,~4$ items); \emph{Behavioral Intention} ($\omega = 0.916,~3$ items).~
Participant-level UTAUT construct composites were significantly above the neutral midpoint in two-sided Wilcoxon signed-rank tests with Holm correction, with large rank-biserial effect sizes: \emph{Performance Expectancy} ($M=0.78$, $Mdn=0.84$, $W=9.0$, $p_{\mathrm{Holm}}<0.001$, $r_{\mathrm{rb}}=0.96$), \emph{Effort Expectancy} ($M=0.83$, $Mdn=0.84$, $W=2.0$, $p_{\mathrm{Holm}}<0.001$, $r_{\mathrm{rb}}=0.99$), \emph{Facilitating Conditions} ($M=0.74$, $Mdn=0.75$, $W=1.5$, $p_{\mathrm{Holm}}<0.001$, $r_{\mathrm{rb}}=0.99$), and \emph{Behavioral Intention} ($M=0.80$, $Mdn=0.86$, $W=4.0$, $p_{\mathrm{Holm}}<0.001$, $r_{\mathrm{rb}}=0.97$).
The mean normalized UTAUT score across participants was $M = 0.78$, $SD = 0.12$, $95\%~CI~[0.73,~0.84]$, and the mean \emph{Behavioral Intention} score was $M = 0.80$, $SD = 0.15$, $95\%~CI~[0.73,~0.86]$.~\Cref{fig:boxes} (a) shows boxplots of the UTAUT scores.

Following the UTAUT framework, we fitted multiple linear regressions with and without moderators. However, due to the small sample size, all the regression analyses remain exploratory.
A multiple linear regression without the moderators accounted for 67\% of the variance in \emph{Behavioral Intention} ($R^2 = 0.67$, $R^2_{\text{adj}} = 0.62$, $F = 6.06$, $p < 0.01$).
In this model, \emph{Performance Expectancy} showed the strongest association with \emph{Behavioral Intention} ($\beta = 0.67,~p = 0.014$), whereas \emph{Facilitating Conditions} ($\beta = 0.10,~p = 0.48$) and \emph{Effort Expectancy} ($\beta = 0.12,~p = 0.48$) were not significant.
A multiple linear regression including age, gender, and expertise as moderators accounted for 83\% of the variance in \emph{Behavioral Intention}, but a substantially lower $R^2_{\text{adj}}$ ($R^2 = 0.83$, $R^2_{\text{adj}} = 0.57,~F = 2.83,~p = 0.060$) and did not show reliable associations with \emph{Behavioral Intention}.

\subsubsection{HCI Toolkit Goals}

Some constructs of our customized HCITG scale showed acceptable reliability ($\omega \geq 0.7$): \emph{Enabling Creative Exploration} ($\omega = 0.830,~3$ items); \emph{Empowering New Audiences} ($\omega = 0.828,~5$ items); \emph{Integrating with Current Practices and Infrastructure} ($\omega = 0.786,~3$ items); \emph{Enabling Replication of Existing Solutions} ($\omega = 0.784,~3$ items).
However, others fell below this threshold: \emph{Reducing Authoring Time} ($\omega = 0.687,~3$ items); \emph{Creating Paths of Least Resistance} ($\omega = 0.655,~3$ items).
Participant-level HCITG construct composites were significantly above the neutral midpoint in two-sided Wilcoxon signed-rank tests with Holm correction. All effects were large by rank-biserial correlation: \emph{Reducing Authoring Time} ($M=0.81$, $Mdn=0.81$, $W=1.0$, $p_{\mathrm{Holm}}<0.001$, $r_{\mathrm{rb}}=0.99$), \emph{Creating Paths of Least Resistance} ($M=0.68$, $Mdn=0.67$, $W=6.0$, $p_{\mathrm{Holm}}<0.001$, $r_{\mathrm{rb}}=0.96$), \emph{Empowering New Audiences} ($M=0.78$, $Mdn=0.81$, $W=1.0$, $p_{\mathrm{Holm}}<0.001$, $r_{\mathrm{rb}}=0.99$), \emph{Integrating with Current Practices and Infrastructure} ($M=0.72$, $Mdn=0.69$, $W=8.0$, $p_{\mathrm{Holm}}<0.001$, $r_{\mathrm{rb}}=0.95$), \emph{Enabling Replication of Existing Solutions} ($M=0.76$, $Mdn=0.76$, $W=1.0$, $p_{\mathrm{Holm}}<0.001$, $r_{\mathrm{rb}}=0.99$), and \emph{Enabling Creative Exploration} ($M=0.78$, $Mdn=0.79$, $W=4.0$, $p_{\mathrm{Holm}}<0.001$, $r_{\mathrm{rb}}=0.97$).
The mean HCITG score across participants was $M = 0.75$, $SD = 0.09$, $95\%~CI~[0.71,~0.79]$.
~\Cref{fig:boxes} (b) shows boxplots of the HCITG scores.

We examined marginal associations between the HCITG constructs and \emph{Behavioral Intention}. All six constructs showed positive Pearson and Spearman correlations with \emph{Behavioral Intention}. The strongest Pearson correlations were observed for \emph{Reducing Authoring Time} ($r=0.59$), \emph{Enabling Replication of Existing Solutions} ($r=0.57$), \emph{Enabling Creative Exploration} ($r=0.55$), and \emph{Integrating with Current Practices and Infrastructure} ($r=0.55$). The strongest Spearman correlations were observed for \emph{Creating Paths of Least Resistance} ($\rho=0.52$) and \emph{Integrating with Current Practices and Infrastructure} ($\rho=0.52$), followed by \emph{Enabling Replication of Existing Solutions} ($\rho=0.42$) and \emph{Reducing Authoring Time} ($\rho=0.36$). Overall, the correlations suggest that all HCITG constructs were positively associated with \emph{Behavioral Intention}, with \emph{Reducing Authoring Time} showing the strongest linear association and \emph{Creating Paths of Least Resistance} and \emph{Current Practices} showing the strongest monotonic associations.

\subsubsection{Toolkit Comprehension}

The average \emph{comprehension} score across participants was $M = 0.92$, $SD = 0.09$, $95\%~CI~[0.88,~0.96]$.
The numerically highest walkthrough comprehension score was obtained on the tasks \textit{Toolkit Overview} ($M = 0.97,~SD = 0.13$) and \textit{Setup} ($M = 0.97,~SD = 0.09$).
The numerically lowest video walkthrough comprehension score was obtained on the tasks \textit{MCP Integration} ($M = 0.83,~SD = 0.28$) and \textit{Modifying XARP} ($M = 0.85$, $SD = 0.27$).
The \emph{self-rated confidence on UTAUT answers} average across participants was $M = 0.80$, $SD = 0.15$, $95\%~CI~[0.73,~0.87]$.
The \emph{self-rated confidence on HCITG answers} average across participants was $M = 0.80$, $SD = 0.12$, $95\%~CI~[0.75,~0.86]$.

A rapid Framework analysis~\cite{gale2013method} of participant responses to the question, ``What other questions do you have that were not answered in the videos?'' produced seven main FAQ topics. The most frequent topic was \emph{Compatibility \& Integration} (42\%), e.g., ``Is XARP compatible with other AR headsets beyond the Quest?'' and ``Does the [toolkit] have anything for integrating with Unity Sentis and on-device AI models?'' Another frequent topic was \emph{Limitations} (15\%), e.g., ``(...) In what circumstances have you found the built-in commands, LLM prompting, or motion/data tracking to be unreliable?''

\subsubsection{Qualitative Basis for Behavioral Intention}

Consistent with the theoretical structure of UTAUT, our regression analysis suggests that \emph{Behavioral Intention} was associated with \emph{Performance Expectancy}. Qualitative accounts corroborate this finding and further reveal reasons aligned with \emph{Reducing Authoring Time}, with participants identifying three main sources of perceived gains:

\begin{itemize}
    \item \textbf{High-level abstractions}: Participants valued the simplicity and clarity of XARP’s functions and its ability to hide unnecessary technical detail. P3 highlighted ``organization/simplicity of commands, and ease of integration with external tools,'' adding that they prefer to ``use Python when possible.'' P7 commented that XARP can ``expose the least low-level details to the users who are not experts in AI or XR.'' P9 expected that they ``would not get bogged down with reading up on various concepts or worrying about smaller details'' when using XARP.
    \item \textbf{Live reloading}: Several participants associated performance gains with the ability to code server-side apps without redeploying clients. P2 remarked that with XARP, there is ``no need to build and rebuild Unity Projects.'' Similarly, P24 appreciated that XARP ``allows developers to avoid constantly pushing new builds in the development of an XR app.''
    \item \textbf{AI agent integration}: Participants also saw performance benefits in XARP’s AI agent integration. P6 valued how ``the agent understands my need and my current status.'' P17 emphasized time savings for prototyping: ``the main value add would be the amount of time it would save to come up with quick prototypes (...) I think a strength is the flexibility to swap out models and plug-and-play MCPs.''
\end{itemize}

When questioned about reasons why they might avoid using XARP, participants identified reasons aligned with \emph{Facilitating Conditions}:

\begin{itemize}
    \item \textbf{Compatibility}: Participants expressed concerns about device and client support. P2 noted ``Client compatibility with different headsets,'' while P4 remarked, ``I probably would not use it in my current role, because I do not have access to a Meta Quest device.'' Similarly, P24 emphasized, ``I would not use the platform if I did not own a Meta Quest.'' These concerns highlight that the toolkit's portability does not benefit users whose target platform we have not yet ported the client to. We currently support Meta Quest and Android ARCore, but new clients can be added through the Unity multiplatform pipeline and inherit compatibility with any XARP server application.
        \item \textbf{Learning Curve and Documentation}: Several participants highlighted the need for more learning support and documentation. P3 commented, ``If there were more fitting alternative technologies in certain circumstances (or technologies with more thorough documentation or other similar benefits), then I might be preferential to those. I don't think I have enough knowledge/info to respond specifically to the downsides of this platform.'' Others pointed to the difficulty of learning and using the system effectively: P6 mentioned ``The sharp learning curve or the long running time it takes,'' P10 said they ``Need more time to understand and learn how to use it and design more functions,'' and P14 noted ``Difficulty in learning the language syntax.'' We plan to create comprehensive documentation for XARP once it is expanded to include the full planned range of functionalities, which we anticipate will ease the concerns we observed regarding learning support.
    \item \textbf{Performance}: Although \emph{Performance Expectancy} was the strongest predictor of \emph{Behavioral Intention}, participants also noted constraints related to computational performance. P7 observed, ``If the token usage is too high, it might stop users from creating more creative or complex tasks, (because) not too many users have a locally deployed GPT-oss model, or they could not afford the cost of using token APIs.'' Likewise, P12 raised concerns about performance and reliability: ``Performance for real-time tasks, not sure how an AI coding agent will be able to optimize an XR app to run in real-time or improve performance. I feel that client-server interaction is unreliable.'' P23 was concerned that the toolkit ``performs I/O and rendering in a (...) potentially non-performant way.''
\end{itemize}

\section{Longitudinal Evaluation}

After incorporating participant feedback from the early acceptance evaluation, we sought opportunities to observe XARP in use on independent research projects. To disseminate the toolkit and prospect early adopters, we conducted a demonstration open to students in the laboratory of one of the authors. Following the demonstration, a researcher not among the authors of this work decided to adopt XARP in their own project, which was originally intended to use Unity as its development platform. This opportunity allowed us to conduct a longitudinal study examining toolkit utilization in an ecologically valid setting, where the objectives, timeline, and personnel were established entirely independently of our evaluation goals.

We followed a team of two researchers over six weeks as they developed their independent research project with XARP. Through repeated checkpoints and a final interview, we captured both initial experiences and evolving perceptions of the toolkit. Our analysis focused on identifying barriers to initial adoption, limitations, and patterns of toolkit appropriation through sustained use~\cite{audulv2023time}. Participants provided informed consent and reviewed this section to ensure anonymization and nondisclosure of details related to their unpublished research. This evaluation was approved by our local institutional review board (UCSB \#39-25-0548).

\subsection{Participants}
The study was conducted with a team of two XR-AI developers. PL1 is an undergraduate in Computer Science (18--24, F) with less than one year of Unity development experience on Meta Quest 3, and AI experience deploying 3D generation models~\cite{sam3dteam2025sam3d3dfyimages} and using AI APIs (Hugging Face, OpenAI). PL2 is a PhD student (18--24, M) with one to three years of XR development across Meta Quest, HTC Vive, and Magic Leap using Unity, and experience training and deploying AI models using tools such as Hugging Face, TensorFlow, and Ollama. Both are close collaborators at our university and not authors of this article.

\subsection{Case Study Overview}

Because the research project used as the case study in this evaluation remains unpublished at the time of writing, we refrain from reporting its motivation or hypotheses. Instead, we focus on how XARP was used in the project. Participants used the toolkit to build a system that enabled users to capture egocentric RGB images through hand gestures. These images are processed by a vision-language model and presented back to the user alongside textual and audio information in a structured layout. We report on six weeks of case study development.

\subsection{Procedure}
Our data collection methodology consisted of impromptu meetings with participants throughout six weeks, with touchpoints adapted to their natural workflow rather than imposed at fixed intervals~\cite{audulv2023time}. On the first day, participants received the Python package, Unity client binary for Meta Quest, and an installation tutorial covering content equivalent to~\cref{sec:how_to}. One author was available during business hours throughout the study to provide technical support when needed. These interactions forced us to refine our documentation, a concern raised in the early adoption study, and shaped the content presented in~\cref{sec:api} and~\cref{sec:arch}. Identified bugs were resolved iteratively, including issues with gesture detection and element hierarchy construction. After six weeks of sustained engagement with the toolkit, the supporting author conducted a one-hour semi-structured interview with participants (script available in the supplemental materials). The interview was transcribed and complemented with field notes, covering topics including workflow evolution, comparisons with Unity, pain points encountered during development, and feature requests, with particular emphasis on initial setup experiences and limitations that emerged through sustained use.

\subsection{Data Analysis}
The same author who supported participants during this study conducted the data analysis. They started by compiling notes of their touchpoints with participants. Interview transcriptions were organized using the interviewer's notes as an initial outline. These data sources were combined in a unified dataset and indexed by week, facilitating the identification of changes in behavior over time~\cite{audulv2023time}.

\subsection{Results}
\subsubsection{Weeks 1--2 - Gaining Momentum}
Both participants found XARP easier to learn than Unity. PL1, who had recently set up an XR project in Unity, called XARP setup \emph{``way easier''}, recalling that their \emph{``first build took 20 minutes''} in Unity, with troubleshooting the headset connection taking \emph{``an hour.''} With XARP, initial deployment happened within minutes. PL2, with more extensive XR experience, observed that XARP bypasses common Unity hurdles such as installing XR packages, resolving version conflicts, and configuring headset connections. Both followed a tutorial similar to~\cref{sec:how_to} and began developing without difficulty.

Participants established a workflow that mirrored their existing Python practices: modify code, run, observe results on the headset, repeat. They debugged using print statements and error messages, with occasional client restarts to ensure a fresh client state. PL2 highlighted the convenience of connecting via QR code, eliminating the need for cables.

\subsubsection{Weeks 3--5 - Limitations \& Workarounds}
As participants tackled more complex development, they encountered limitations addressed through workarounds or feature requests. PL2 requested standard UI elements (buttons, toggles, sliders) and visual effects such as line renderers and hand occlusion, drawing on features available in other XR toolkits. Both participants wanted built-in poke/grab interactions and co-location support. These requests are currently part of our development roadmap.

Access to source code proved valuable: participants extended XARP rather than waiting for fixes. PL2 modified \texttt{xarp.update} to accept lists after discovering a performance bottleneck from single-object updates. They customized the look-and-feel of their interface using Python libraries to generate images displayed as \texttt{ImageAsset}s. The need for more thorough documentation---particularly conceptual guides and task-oriented tutorials---also became apparent as source reading alone failed to scale.

\subsubsection{Week 6 - Reflecting on Sustained Use}
Prolonged headset wear became fatiguing, prompting requests for a desktop simulator for small headset-free tests. Participants reflected on ecosystem tradeoffs. Python's AI ecosystem was a clear asset for their vision-language model integration, while the absence of game engine asset marketplaces was a recognized but non-critical limitation in their project. They identified that asset-intensive or performance-critical projects might be a poor fit for XARP.

Both participants responded affirmatively when asked whether they would choose XARP again, citing easier comprehension, faster startup, better iteration speed, and natural AI integration. Neither wished they had started with Unity despite the limitations encountered. They recommended XARP most strongly for beginners and developers unfamiliar with Unity. PL2 added that experienced developers might benefit from XARP for rapid prototyping.

\section{Agent Performance Benchmark}
\label{sec:agent_eval}

Addressing questions about token consumption raised in our early acceptance evaluation, we conducted a technical benchmark comparing the token consumption of AI agents using our toolkit and conventional C\# Unity in a series of XR application generation tasks. Results indicate XARP enabled a 19\% median reduction in token consumption.

\begin{figure}
    \centering
    \includegraphics[width=.8\linewidth]{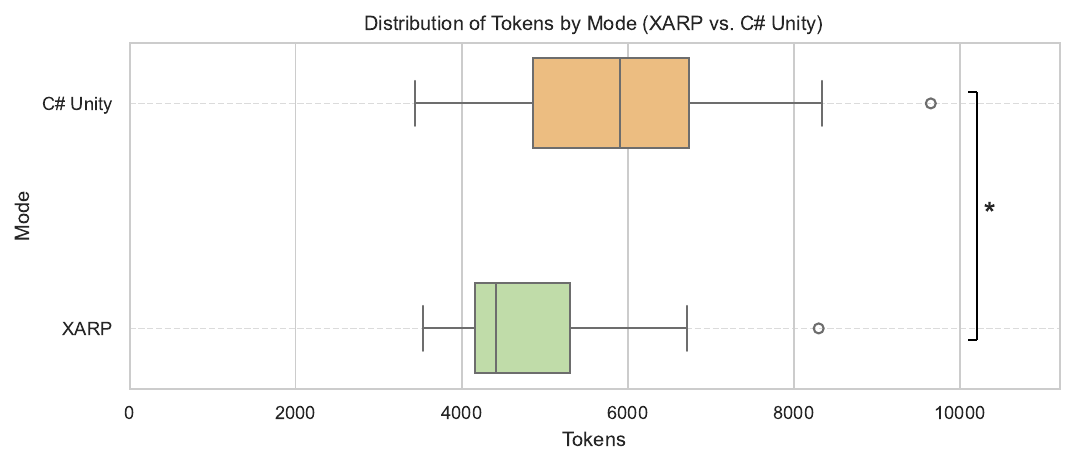}
    \caption{Token consumption of AI agents for the baseline C\#/Unity compared to XARP. Agents consumed significantly more tokens in the baseline condition than with XARP. The figure is truncated at 10,000 tokens for clarity, and excludes one outlier in each condition.}
    \label{fig:tokens}
    \Description{Figure showing horizontal box plots comparing AI agent token consumption between two XR development workflows: XARP and a baseline C\#/Unity workflow. The x-axis represents token counts, and the y-axis lists the two development modes. The XARP condition shows a lower token distribution overall, with most values clustered roughly between 4,000 and 5,500 tokens and a smaller interquartile range. The C\#/Unity condition shows consistently higher token consumption, with most values distributed roughly between 5,000 and 7,000 tokens and a higher upper range approaching 10,000 tokens. Each condition contains an outlier beyond the main whisker range, though the figure truncates values at 10,000 tokens for readability. A significance marker indicates a statistically significant difference between the two conditions, with the baseline C\#/Unity workflow requiring substantially more tokens than XARP.}
\end{figure}

\subsection{Apparatus}

We implemented two autonomous code-generation agents using Devstral Small 2 and the Hugging Face Smolagents framework. Devstral Small 2 was selected for its strong coding performance relative to parameter count~\cite{mistral2025devstral2}, vision capability, open-weight availability, and training on tool use, making it suitable for local execution and runtime integration with XARP.

One agent generated XR applications in Python using XARP, while the baseline agent generated functionally equivalent applications in C\# for Unity.  The context window size was fixed at 40,960 tokens, and the sampling temperature was set to 0.7. All experiments were executed on a single NVIDIA RTX 3090 GPU. Agents were implemented as XARP XR agents as seen in~\cref{sec:how_to_agents}. To avoid network noise, the XARP client ran in the Unity editor during the experiments.

\subsection{Tasks}

We compared conditions across 14 tasks, eight designed by us and six taken from prior work~\cite{de2024llmr}. Tasks require interactive behavior and the manipulation of primitive geometries. For example, spawning a sphere when the user enters a cubic trigger volume and animating a grid of spheres using a time-varying sinusoidal function. All tasks were constructed to be solvable in both conditions and to require comparable levels of control flow, spatial reasoning, and state management. Tasks are available in the supplemental material.

\subsection{Procedure}

Each agent completed all 14 tasks independently. For each task, the agent was allowed to iteratively generate code until a correct, functioning application was produced. All trials resulted in successful task completion by design, ensuring that comparisons of token consumption reflected differences in generation efficiency rather than task success or failure. For each trial, we recorded the total number of tokens consumed across the complete interaction required to generate a working XR application.

\subsection{Analysis and Results}

We conducted a paired-samples comparison of token usage across the 14 tasks ($n = 14$ pairs). The baseline condition consumed a mean of 7,000 tokens ($SD = 4,384;~Mdn = 5,908$) per task, while XARP consumed a mean of 5,283 tokens ($SD = 2,199;~Mdn = 4,414$) (see~\cref{fig:tokens}). In 10 of the 14 tasks, XARP required fewer tokens than the baseline. Because paired differences were not normally distributed (Shapiro-Wilk; $SW = 0.86, p = 0.034$), we used the Wilcoxon signed-rank test rather than a paired t-test.
The Wilcoxon signed-rank test supported the directional hypothesis that XARP would consume fewer tokens than the baseline ($W = 21, p = 0.025$, one-tailed). The rank-biserial correlation of $r_{\mathrm{rb}}=0.60$ indicates a large effect~\cite{peres2026effect}. The median token reduction was 1,081 tokens per task, corresponding to a 19\% decrease from baseline.~\cref{fig:tokens} shows the token distributions in both conditions.

\section{Sensing Benchmark}

The early acceptance evaluation raised concerns about I/O performance over the network, as noted by P12 and P23. To address these concerns, we evaluated the sensing performance of XARP by measuring throughput in frames per second (FPS) across combinations of sensing modalities under repeated single-response calls and streaming response mode. Experiments were conducted in two hardware setups. The \emph{local editor} setup consisted of a Unity editor connected to a localhost server, offering a baseline free of network effects. The \emph{on-device} setup ran a Meta Quest 3 connected via a university Wi-Fi network (2 Gbps upload and download) during business hours, approximating typical operational conditions for both network bandwidth and on-device processing. The Meta Quest 3 device operated on a 72 Hz display refresh rate.

\subsection{Procedure}
Each combination of the modalities \texttt{head}, \texttt{hands}, and \texttt{image} was evaluated under both response modes and platforms, yielding $7\times2\times2=28$ configurations in total. For each configuration, three independent runs were performed. In each run, the first 10 frames were discarded as warm-up; the measurement window then consisted of $N = 1000$ consecutive frame intervals. Timestamps were recorded immediately after a data frame was decoded and available to the XARP application. Throughput was computed as $\mathrm{FPS} = N / (t_N - t_0)$, where $t_0$ and $t_N$ are the timestamps of the first and last measured frames, respectively.
All reported values are the mean across the three runs. The procedure was identical for both the local editor and on-device setups.

\subsection{Results}
\begin{figure}[t]
  \centering
    \includegraphics[width=\linewidth, trim={0 1cm 0 0},clip]{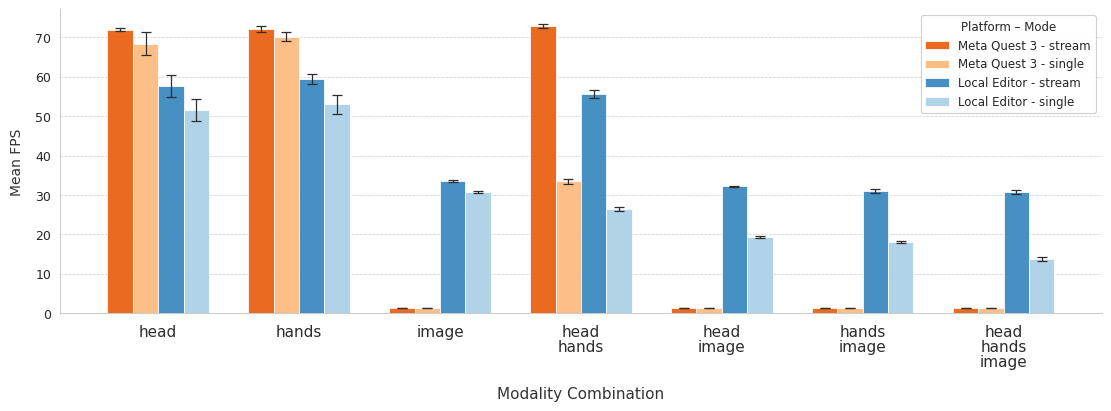}
  \caption{Throughput performance across modality combinations and execution modes (FPS, higher is better). Stream mode consistently outperforms single-response mode. Image capture shows substantial hardware dependency, with throughput dropping from 33.6 FPS in the local editor to 1.4 FPS on Meta Quest 3.}
  \label{fig:fps}
\Description{Bar chart comparing throughput performance in frames per second (FPS) across different sensing modality combinations and execution modes in XARP. The x-axis lists modality configurations including head tracking, hand tracking, image capture, combined head and hands, combined head and image, combined hands and image, and combined head, hands, and image. The y-axis shows mean FPS. Four execution configurations are compared using colored bars: Meta Quest 3 stream mode, Meta Quest 3 single-response mode, local editor stream mode, and local editor single-response mode. For head-only and hands-only sensing, Meta Quest 3 stream mode achieves approximately 72 FPS, while local editor stream mode achieves around 58 to 60 FPS. Image capture exhibits much lower performance on Meta Quest 3, around 1 to 2 FPS, but substantially higher performance in the local editor, around 34 FPS in stream mode. Combining image capture with head or hand tracking further reduces FPS, especially on Meta Quest 3. Across all modality combinations, stream mode consistently outperforms single-response mode. The figure demonstrates that image acquisition is the primary performance bottleneck and that throughput strongly depends on both execution mode and hardware platform.}
\end{figure}

Throughput (FPS; higher is better) was higher in the stream response mode across data modality combinations and hardware conditions. The streaming advantage compared to repeated single responses was most pronounced when reading \texttt{head} and \texttt{hands} together on Meta Quest 3, achieving 66.7 FPS while single-response mode dropped to 33.4 FPS, representing a 50\% reduction in throughput when switching from streaming to single responses.
In the local editor, \texttt{image} streaming reached 33.6 FPS, but there was a drastic drop on Meta Quest 3 to 1.36 FPS. Any modality combination including \texttt{image} was bounded by this rate on Meta Quest 3, ranging between 1.31 and 1.42 FPS regardless of mode or additional modalities.
The highest measured throughput was 72.1 FPS for streaming \texttt{hands} data on Meta Quest 3 compared to 59.4 FPS on the local editor.~\Cref{fig:fps} shows the mean throughput for each configuration.

\section{Discussion}

\subsection{Lowering Barriers in XR Development}
The main goal of XARP is to broaden access to XR development, which is currently limited to those familiar with game engines and C\#/C++ coding. By exposing XR operations through a Python library, XARP allows Python developers, including a considerable number of AI researchers, to create XR interfaces using their familiar workflows. The example applications in~\Crefrange{sec:how_to}{sec:how_to_mcp} demonstrate how XARP effectively enables XR development in Python. Our early acceptance study showed that XR and AI developers expected XARP to reduce their effort and accelerate their development process. The XR abstraction available in the toolkit, the decoupling of XR device and application logic, and the live reloading feature all contributed to this early acceptance. These patterns were observed again after sustained six-week usage as reported in our longitudinal study.

The expected benefits of XARP extend beyond novices to experienced XR developers who can use XARP for rapid prototyping. The emerging MCP ecosystem offers a common integration layer to quickly compose off-the-shelf AI capabilities for no-code prototyping. The ability to delegate parts of application behavior to AI agents who can make decisions and use XR capabilities through XARP represents another emerging design approach to tackle requirements uncertainty at runtime.

\subsection{Toolkit Ceiling}

The formative and evaluative processes revealed both temporary and strict ceilings~\cite{myers2000past} of XARP. The temporary ceiling encompasses missing features that can be addressed through planned development, while the strict ceiling limitations on what the toolkit can achieve are derived from core design tradeoffs.

\subsubsection{Temporary Ceiling: Roadmap Items}
Sustained use in the longitudinal study identified a roadmap of features that are technically feasible but not yet implemented. Participant requests included standard UI widgets, visual effects such as line renderers, and particles. Similarly, the toolkit does not yet support lighting, shader authoring, particle systems, and animations. The need for more comprehensive documentation also became apparent. These temporary limitations reflect the current maturity level of a research prototype rather than fundamental constraints. In the timespan of this article, several limitations were addressed, for example, the inclusion of stream response mode that addresses concerns about I/O efficiency identified in the early acceptance evaluation.

\subsubsection{Strict Ceiling: Architectural Limitations}
Our evaluations revealed use cases that have a poor fit with XARP due to 
core toolkit design decisions. While participants acknowledged faster on-device iteration compared to traditional XR workflows, prolonged headset use became tiring, leading them to request a desktop simulator for headset-free testing. Participants also reflected on ecosystem tradeoffs. Gaining access to the Python ecosystem for AI integration meant losing access to game engine asset marketplaces and fine-grained control over rendering. As a result, asset-intensive projects might not benefit from XARP. Similarly, performance-critical applications requiring lower-level optimizations are poor fits for the toolkit. Composing large numbers of assets without a visual editor becomes challenging.

Sensing benchmarks revealed a throughput gap between the local editor and the Meta Quest 3 during image streaming, likely caused by images being transmitted as uncompressed 1280x720 bitmaps. Adding JPG compression to captured images brought frame rates above 30 FPS in a preliminary image streaming performance test, although thorough benchmarking remains future work. XARP's streaming mechanism was designed for small payloads and lacks throttling or compression. For long-term multimedia streaming, alternatives such as WebRTC may be considered.

\subsection{Design Reflection}
Throughout the development of this toolkit, we identified a series of design tradeoffs that may be relevant for other researchers and interactive system engineers. We describe those tradeoffs and offer possible alternatives when applicable.

\subsubsection{Human-First vs. Agent-Ready}
Since the initial stages, XARP has prioritized keeping a human-friendly public API while allowing AI agents to share that same API. As we implemented examples and benchmarks with AI agents and MCP, we noticed a tension between these goals. Rich human-oriented APIs rely on complex data models such as \texttt{Element} and implicit conventions such as iterating over a sense stream as the main application loop. Agent-oriented APIs favor simpler primitives with less polymorphic behavior, in line with best practices for callable tools. MCP is still maturing, and its evolving patterns introduce additional pressure. Current reference implementations such as FastMCP\footnote{FastMCP: \url{https://gofastmcp.com}. Accessed March 27, 2026.} may require MCP-specific data types such as \texttt{MCPImage}\footnote{MCPImage FastMCP: \url{https://gofastmcp.com/python-sdk/fastmcp-utilities-types}. Accessed March 27, 2026.} that are not ergonomic for human use. Allowing fast-changing third-party interfaces to leak into the toolkit risks eroding the Human-First principle and undermining XARP's primary utility as a Python-based XR development platform. Our strategy is to monitor how MCP APIs and patterns evolve before committing to deeper integration. Looking further ahead, as AI coding agents continue to improve, they may reach a level where they can consume human-oriented APIs as naturally as human developers do, at which point the distinction between human-first and agent-ready design may disappear.

\subsubsection{Portability}

XARP's portability rests on the separation between server-side application logic and the platform-specific client. The client is implemented in Unity, with prebuilt binaries currently available for Meta Quest and Android ARCore. Supporting a new platform requires porting only the client, a manageable task given Unity and OpenXR's broad device coverage; Magic Leap, Pico, and HoloLens are all candidates. Once ported, all existing XARP applications become compatible with that platform without further modification.

This strategy offers a considerable portability range, but has limitations. Porting to a less capable device, such as one without 6DOF head tracking, leaves the corresponding API methods present but non-functional at the device level, pulling API design toward the lowest common denominator of all target platforms. The symmetric pressure occurs when a target platform exposes a unique capability absent from the XARP API. Including it breaks compatibility with devices that lack that feature and leads to API bloat. Platforms without a Unity pipeline introduce a third, independent limitation. Supporting them requires a full client reimplementation, and only the Python server code carries over. Implementation differences between the original and the reimplemented client may produce behavioral divergence that is difficult to anticipate.

The right approach to managing this tension is not yet settled. Two plausible models are OpenXR's extension mechanism, where applications query which extensions the underlying runtime supports at initialization and enable only a compatible subset, and AR Foundation's subsystem descriptor model, which maps capabilities to platforms explicitly so that developers can query feature availability at startup and implement their own fallback logic accordingly.

\subsubsection{Development Mental Model}
The server-client split requires experienced XR developers to adjust their mental model in several ways. XARP Elements resemble Unity GameObjects and are, in fact, implemented as such on the client, but their semantics differ. \texttt{xarp.update} applies state deltas rather than replacing full entity state, and each update triggers a network call rather than a direct state mutation rendered in the next frame. Update frequency, therefore, carries a cost that developers must account for explicitly.
A related shift concerns the application loop. Unity hides per-frame logic behind lifecycle methods such as \texttt{GameObject.Update}. In XARP, an application may run sequentially or structure its main loop explicitly around a sense stream, keeping control flow visible and central. A third difference is that XARP does not impose an entity-component architecture, leaving developers free to structure their codebase as it grows.
For novice XR developers with no prior Unity experience, none of these are adjustments. They are simply defaults, which may be one reason novices might be more willing to use XARP.

Token count is a standard efficiency metric in the agent evaluation literature, used to quantify computational cost and verbosity per task. Because tokenizers differ across models, raw counts are not directly comparable between them. Within a single model, however, token count is consistent and reproducible, and this within-model comparison is the basis of our benchmark.

\section{Limitations}

\paragraph{Internal Validity}
The acceptance evaluation mirrors how developers approach a new tool, but the absence of direct usage constrains the data to self-reported perceptions rather than observed behavior. The sample (n=24) is small, making most of its analysis exploratory. The participants were nonetheless diverse and representative of the target audience, spanning developers and researchers with varying XR backgrounds. The longitudinal study was conducted by researchers independent from this paper, though institutional proximity and experimenter expectation effects may still have influenced outcomes. The study ran for six weeks and included both a novice and an expert participant, covering a meaningful range of experience levels over a non-trivial duration.

\paragraph{External Validity}
Technical evaluations were performed with a Meta Quest 3 and Devstral Small 2 on a university network, and results may differ with other devices, models, or deployment conditions. Both configurations are nonetheless accessible and widely used, which strengthens the practical relevance of the findings. The early acceptance study relied on recruitment through the authors' social media channels, which may have introduced self-selection bias toward participants already familiar with or favorably disposed to the work. The longitudinal participants were drawn from a single institution, which may limit generalizability to professional developers or broader industry contexts.

\paragraph{Construct Validity}
Social Influence was excluded from the UTAUT model because the study targeted individual early acceptance following a walkthrough session rather than organizational adoption, making the factor conceptually inappropriate rather than merely inconvenient. The exclusion nonetheless limits direct comparability with prior UTAUT studies. Token count was used as a proxy for agent efficiency, a standard measure in the agent evaluation literature for capturing verbosity and computational cost per task. It is internally consistent when comparing runs of the same model, though it does not generalize across models due to differences in tokenization. The HCITG instrument was a customized survey without prior psychometric validation. It demonstrated reasonable internal consistency (McDonald's omega above 0.7) and included reverse-coded items to reduce acquiescence bias.

\section{Future Work}
In addition to implementing the prioritized missing features gathered throughout the development and evaluation process of XARP, we are currently working on several directions that extend the toolkit's impact and sustainability.

\paragraph{Game Engine Interoperability}
Although XARP offers a self-contained development workflow, we plan to support interoperability with game engines. The primary motivation is access to engine-level assets, XR toolkits, and legacy projects not covered by the XARP API. Because XARP is built on Unity and OpenXR, the natural first step is integrating the XARP client into existing Unity projects, enabling gradual migration or coexistence of Unity and XARP components within the same application. This would address an ecosystem tradeoff identified in our evaluations, allowing developers to draw on rich game engine assets while benefiting from XARP's rapid prototyping workflow.

\paragraph{Open-Source Community} 
XARP is an open-source project, with both the Unity client and the Python backend library publicly available. Community contributions are central to its long-term trajectory. They accelerate roadmap delivery, surface diverse use cases, expand platform support, and keep the toolkit current with advances in AI and XR hardware. Broader participation also improves documentation grounded in real-world usage, and extends the reach of XARP's core goal of lowering the barrier to XR development.

\paragraph{Supporting Spatial Human-AI Interaction Research}
Beyond lowering barriers to XR development, XARP has broader implications for spatial human-AI interaction research. A shared abstraction layer for XR accelerates hypothesis testing through faster prototyping and improves the replicability of research systems, as studies built on XARP can be reproduced across platforms without reimplementing device-specific infrastructure. The multiplatform architecture also supports more rigorous experimental methodology by enabling systematic comparison of design alternatives across devices. As new XR hardware capabilities are integrated into the library, the research community gains immediate access to them without rebuilding low-level integrations. Because XARP decouples research systems from any single vendor's SDK or commercial roadmap, studies remain reproducible even as device ecosystems evolve or specific platforms are discontinued. XARP additionally brings AI researchers into the spatial computing community, potentially catalyzing novel interaction paradigms. Its agent-ready design positions the toolkit as a platform for exploring human-AI collaboration in spatial contexts, from runtime behavior generation to automated XR application testing.

\section{Conclusion}

We presented XARP, a Python toolkit for prototyping XR-AI systems. Application logic runs on a Python server that controls a Unity client over WebSocket, bypassing game engines and supporting multiple devices through a single client port. The same abstraction is exposed to AI agents as callable tools and through the Model Context Protocol. An acceptance survey with 24 developers and a six-week longitudinal deployment indicated faster setup and iteration than conventional XR workflows, while identifying asset-intensive and performance-critical projects as the toolkit's ceiling. In benchmarks, AI agents using XARP consumed 19\% fewer tokens at the median than agents producing equivalent C\#/Unity code. XARP is open source and available at \url{https://github.com/hal-ucsb/xarp}.

\begin{acks}
We thank the U.S. National Science Foundation for supporting this research through the Early CAREER Award 2023 no. 2240133.
We especially thank Yunhao Luo, Irene Li, Avinash Nargund, Celine Zhao, and the students of the UCSB Winter 2026 CS291I class, whose hands-on engagement and feedback helped shape XARP.
\end{acks}

\bibliographystyle{ACM-Reference-Format}
\bibliography{main}
\end{document}